\documentclass[11pt]{article}
\pdfoutput=1

\usepackage{jheppub} 

\usepackage{amsmath,amssymb,epsfig,amsfonts}
\usepackage{graphicx,subfigure}

\usepackage{epsf}
\usepackage{makeidx}
\usepackage[all]{xy}
\usepackage{slashed}
\usepackage{verbatim} 
\usepackage{float}
\restylefloat{table}
\usepackage[all]{xy}
\usepackage{pdflscape}
\usepackage{tikz}
\usepackage{array}
\usepackage{multirow}
\usepackage{color}
\usepackage{ulem}
\usepackage{cleveref}
\usepackage{mathrsfs}
\usepackage{tensor}
\usepackage{rotating}

\makeatletter

\newcommand{\be}{\begin{equation}}
\newcommand{\ee}{\end{equation}}
\newcommand{\dd}{\mathrm{d}}
\newcommand{\me}{\mathrm{e}}

\newcommand{\vol}{\mathrm{vol}}

\newcommand{\ls}{\ell_{\rm s}}

\newcommand{\tr}{\mathrm{Tr}}

\makeatother

\begin{document}

\baselineskip=18pt  
\numberwithin{equation}{section}  
\allowdisplaybreaks  


\thispagestyle{empty}

\begin{titlepage}

\vspace*{1cm} 
\begin{center}
{\Large \bf  $\mathcal{N}=(0,4)$ Black String Chains} 
 \vskip 1cm

{Christopher Couzens$^{\, \mathrm{a}}$, Yolanda Lozano$^{\, \mathrm{b,c}}$, Nicol\`o Petri$^{\, \mathrm{b,c}}$ and Stefan Vandoren$^{\, \mathrm{a}}$}\\

\vskip 0.5cm

${}^{\mathrm{a}}$\textit{Institute for Theoretical Physics and Center for Extreme\\ Matter
and Emergent Phenomena, Utrecht University,\\
Princetonplein 5, 3584 CE Utrecht, The Netherlands\\}

\vskip 0.2cm

${}^{\mathrm{b}}$\textit{Department of Physics, University of Oviedo,\\
Avda. Federico Garcia Lorca s/n, 33007 Oviedo, Spain}

\vskip 0.2cm

${}^{\mathrm{c}}$\textit{Instituto Universitario de Ciencias y Tecnolog\'{\i}as Espaciales de Asturias (ICTEA), \\
Calle de la Independencia 13, 33004 Oviedo, Spain}

\end{center}

\vskip 1 cm

\begin{abstract}

\begin{center}
{\bf Abstract}\

\end{center}

\noindent We construct black string solutions in Type IIA supergravity arising from intersecting D2-D4-D6-NS5 branes in the presence of fractional D4-branes. The fractional D4-branes arise from D6-branes wrapping (collapsing) two-cycles in a Calabi--Yau two-fold. In the near horizon limit these solutions give rise to AdS$_3$ geometries preserving $\mathcal{N}=(0,4)$ supersymmetry and fall within the recent classification of \cite{Lozano:2019emq}. We interpret our solutions as describing chains of black strings stacked on top of each other along an interval. We construct 2d quiver CFTs dual to our solutions that reproduce the Bekenstein-Hawking entropy microscopically.

\end{abstract}

\end{titlepage}
\newpage

\tableofcontents


\section{Introduction}

One of the fundamental problems in theoretical physics is to understand the microscopic origin of the Bekenstein--Hawking entropy. For extremal black holes, which admit an AdS near-horizon limit, AdS/CFT plays an important role.  
The first success in this direction was in \cite{Strominger:1996sh} where five dimensional black holes preserving 16 supersymmetries were studied. These black holes originate from the D1-D5 system in Type IIB compactifications, and occur as black string solutions of 6d supergravity, whose infrared dynamics are described by 2d $\mathcal{N}=(4,4)$ CFTs. 
Later, black holes preserving 8 supersymmetries were constructed in M-theory compactifications on non-singular compact Calabi--Yau threefolds ($\text{CY}_3$) \cite{Maldacena:1997de,Vafa:1997gr,Minasian:1999qn,Castro:2008ne}. Microscopically, they are described by M5 branes wrapped on $\mathbb{R}\times S^1$ times a 4-cycle inside the $\text{CY}_3$, giving rise to strings wrapping $\mathbb{R}\times S^1$, whose infrared dynamics are described by 2d $\mathcal{N}=(0,4)$ CFTs. 
More general bound states of strings with infrared dynamics described by 2d $\mathcal{N}=(0,4)$ CFTs, some of which are of quiver type, have been obtained more recently from M- and F-theory  constructions, see e.g. \cite{Vafa:1997gr,Haghighat:2013tka,Gadde:2015tra,Haghighat:2015ega,Lawrie:2016axq,Couzens:2017way,Couzens:2019wls,Bena:2006qm,Couzens:2020aat}.

In M-theory the interacting strings arise as self-dual strings on the tensor branch of M5-branes probing A- or D-type singularities, or ``end of the space'' M9-branes.  The six dimensional theory living on the M5-branes admits a deformation away from the conformal fixed point where the M5-branes are separated in the extra transverse direction. In this deformation the interacting strings arise as the boundaries of M2-branes stretched between parallel M5-branes. These M2-M5 brane intersections are Hanany--Witten brane set-ups, that support 2d quiver gauge theories. In the IR these give rise to 2d $\mathcal{N}=(0,4)$ CFTs living on self-dual strings in the world-volume of the M5-branes. Given the M-theory origin of these interacting strings they are commonly referred to as M-strings \cite{Haghighat:2013gba}. For M5-branes probing A-type singularities, the case most related to our work in this paper, they support quiver gauge theories with unitary gauge groups. More general quivers involving symplectic, orthogonal and exceptional gauge groups can be obtained from the M-strings associated to M5-branes probing D-type singularities or end of the space M9-branes \cite{Gadde:2015tra}. Given the quiver gauge theories, quantities such as the elliptic genus have been computed using localisation \cite{Lockhart:2012vp,Kim:2012qf,Haghighat:2013gba,Haghighat:2013tka,Gadde:2015tra}.

In \cite{Lozano:2019emq} AdS$_3$ solutions of massive Type IIA preserving $\mathcal{N}=(0,4)$ supersymmetry were classified.\footnote{See also \cite{Martelli:2003ki,Tsimpis:2005kj,Kim:2005ez,Kim:2007hv,Figueras:2007cn,Donos:2008hd,OColgain:2010wlk,DHoker:2008lup,Estes:2012vm,Bachas:2013vza,Benini:2013cda,Jeong:2014iva,Lozano:2015bra,Benini:2015bwz,Kelekci:2016uqv,Couzens:2017way,Eberhardt:2017uup,Dibitetto:2018iar,Dibitetto:2018ftj,Macpherson:2018mif,Legramandi:2019xqd,Lozano:2019jza,Lozano:2019zvg,Lozano:2019ywa,Couzens:2019mkh,Couzens:2019iog,Passias:2019rga,Filippas:2019ihy,Speziali:2019uzn,Lozano:2020bxo,Farakos:2020phe,Couzens:2020aat,Rigatos:2020igd,Faedo:2020nol,Dibitetto:2020bsh,Filippas:2020qku,Passias:2020ubv,Faedo:2020lyw,Eloy:2020uix,Legramandi:2020txf,Zacarias:2021pfz,Emelin:2021gzx,Couzens:2019wls,Couzens:2021tnv,Suh:2021ifj,Boido:2021szx,Ferrero:2020laf,Hosseini:2021fge} for other examples of AdS$_3$ solutions in various supergravity theories.} These solutions include black string near horizons dual to $\mathcal{N}=(0,4)$ quiver CFTs \cite{Lozano:2019zvg}. Moreover, they arise within controlled string theory set-ups with known holographic duals, where the AdS/CFT dictionary can be used. The $\mathcal{N}=(0,4)$ quiver gauge theories constructed in \cite{Lozano:2019zvg} contain two families of unitary gauge groups, $\prod_{i=1}^n$U$(k_i)\times $U$({\tilde k}_i)$. The gauge group U$(k_i)$ is associated to $k_i$ D2-branes while the gauge group U$({\tilde k}_i)$ is associated to ${\tilde k}_i$ D6-branes, wrapped on a compact $\text{CY}_2$. Both D2 and D6 branes are stretched between NS5-branes, in generalised Hanany--Witten  brane set-ups containing two types of colour branes. In addition, D4 and D8 perpendicular branes provide flavour groups to both types of gauge groups, rendering the field theory anomaly-free.

Besides extending the known explicit AdS$_3$ solutions, we give a black string interpretation for these solutions. The black string lives in the asymptotically flat\footnote{Recently there has been much interest in black string solutions which are asymptotically AdS, see for example \cite{Bernamonti:2007bu,Hosseini:2016cyf,Azzola:2018sld,Hosseini:2019lkt,Hosseini:2020vgl,Boido:2021szx,Ferrero:2020laf,Hosseini:2021fge}. In contrast the black strings that we consider here are asymptotically flat.}  background geometry $\mathbb{R}^{1,1}\times \mathbb{R}^3 \times \text{CY}_{2}\times I$, with $I$ an interval which foliates the space.\footnote{In \cite{Lozano:2019emq}  a second class of solutions where the $\text{CY}_2$ is replaced by a general K\"ahler manifold was also constructed.} The black strings arise from a D2-D4-D6-NS5 brane intersection and have near-horizon geometry $\text{AdS}_3\times S^2\times \text{CY}_2\times I$.
\footnote{In \cite{Faedo:2020nol} brane solutions were constructed leading to this type of backgrounds with the compact Calabi--Yau replaced by $\mathbb{R}^4$. This was key to finding an interpretation of these solutions as defects within 5d SCFTs (see also \cite{Faedo:2020lyw}). In order to find a defect interpretation, a non-compact coordinate must become part of the higher dimensional AdS background, AdS$_6$ in this case, in which the defects are embedded. However, the non-compactness of the Calabi--Yau leads to an infinite value of the central charge, obscuring their black string interpretation.} One may further extend this to include D8-branes. The asymptotic geometry is no longer flat but instead conformally flat. 
Our goal in this paper will be to find the black string solutions that lead to the near-horizon geometries constructed in \cite{Lozano:2019emq} in massless type IIA, computing their entropy as well as other observables. We leave the massive extension for future work.  

We begin in section \ref{blackstrings} with the construction of black strings in massless Type IIA arising from intersecting D2-D4-D6-NS5 branes. A key addition, with respect to the brane set-ups discussed in \cite{Lozano:2019emq,Lozano:2019zvg}, will be the inclusion of fractional D4-branes, arising from D6-branes wrapping collapsing two-cycles in the Calabi--Yau. These branes are linked to a closed two form field $H_2$ living on the Calabi--Yau, present in the solutions of \cite{Lozano:2019emq}. However this two-form was set to zero in the global and field theory analysis performed in \cite{Lozano:2019zvg}. In section \ref{blackstringsglobal} we proceed with the global analysis of these solutions, emphasising the new features that arise due to the presence of the $H_2$ form. We quantise the fluxes, identify the source branes of the geometry and compute the holographic central charge. In section \ref{sec:fieldtheory} we turn to the field theory interpretation of these solutions. We construct $\mathcal{N}=(0,4)$ quiver gauge theories that extend the constructions in \cite{Lozano:2019zvg} to include fractional D4-branes. Thus, these quiver CFTs extend further the constructions in \cite{Gadde:2015tra}, once again within well controlled string theory settings with known holographic duals. Our constructions rely on the concrete stringy origin of the multiplets that live and connect the different branes underlying the solutions. We show that they correct the quivers constructed in \cite{Lozano:2019zvg}, not only to include fractional branes but also to modify the fields associated to certain branes. In particular, we show that in our modified quivers there is no need to appeal to a scaling argument, as used in \cite{Lozano:2019zvg} for example, to match to the holographic calculation. 
We conclude in section \ref{conclusions} presenting a number of future directions. Appendix \ref{multiplets} contains a detailed account of the stringy origin of the quivers constructed in section \ref{sec:fieldtheory}. In Appendix \ref{app:Massive} D8-branes are added to this analysis and new quivers in massive Type IIA are constructed that correct those in \cite{Lozano:2019zvg}.


\section{Black strings in massless Type IIA}\label{blackstrings}

In this section we will construct black strings in massless Type IIA arising from intersecting D2-, D4-, D6- and NS5 branes in the presence of fractional D4-branes. The branes are embedded in a $\mathbb{R}^{1,1}\times\mathbb{R}^{3}\times I\times \text{CY}_{2}$ asymptotic geometry and intersected as in table \ref{Table:Dbrane}. The fractional D4-branes arise from D6-branes wrapping (collapsing) two-cycles in the Calabi--Yau. In the following we will consider the fractional branes wrapping a single two-cycle in the Calabi--Yau two-fold, however the generalisation to multiple curves is also possible. We will see that even in the single two-cycle case the analysis is broken up into distinct classes labelled by a positive integer. We begin this section with a general analysis of the brane solution in massless Type IIA supergravity, showing that it is determined by a single differential constraint. We then proceed in taking the near-horizon limit, without specifying an explicit solution to the differential condition, and obtain an AdS$_3$ solution contained in the classification of \cite{Lozano:2019emq}. In section \ref{blackstringsglobal} we proceed in analysing the near-horizon geometry for a given solution of the differential constraint, quantising the fluxes, identifying the source branes of the geometry and computing the central charge. 

\begin{table}
\begin{center}
\begin{tabular}{|c||c|c|c|c|c|c|c|c|c|c|}
\hline
Brane&\multicolumn{2}{c|}{$\mathbb{R}^{1,1}$}&\multicolumn{3}{c|}{$\mathbb{R}^3$}&$I$& \multicolumn{4}{c|}{CY$_2$}\\
\hline
 &  \multicolumn{2}{c|}{}&\multicolumn{3}{c|}{}& & \multicolumn{2}{c|}{$C$}& \multicolumn{2}{c|}{$\bar{C}$}\\
\hline
\hline
D2&$\times$&$\times$&$-$&$-$&$-$&$\times$&$-$&$-$&$-$&$-$\\
\hline
D4& $\times$&$\times$&$\times$&$\times$&$\times$&$-$&$-$&$-$&$-$&$-$\\
\hline
D6&$\times$&$\times$&$-$&$-$&$-$&$\times$&$\times$&$\times$&$\times$&$\times$\\
\hline
D6'&$\times$&$\times$&$\times$&$\times$&$\times$&$-$&$\times$&$\times$&$-$&$-$\\
\hline
NS5& $\times$&$\times$&$-$&$-$&$-$&$-$ &$\times$&$\times$&$\times$&$\times$\\
\hline
NS5'  & $\times$&$\times$&$-$&$\times$&$\times$&$-$ &$\times$&$\times$&$-$&$-$\\
\hline
\end{tabular}
\caption{The brane configuration we will consider. $I$ denotes a line interval of length $2\pi (P+1)$ whilst in the present setup the Calabi--Yau two-fold must be compact and is therefore either $T^4$ or $K3$. $C$ and $\bar{C}$ are dual divisors on the Calabi--Yau two-fold, with the curve $C$ wrapped by the fractional branes.}
\label{Table:Dbrane}
\end{center}
\end{table}


\subsection{Black string solution}

We begin this section by considering a black string arising from intersecting D2-, D4-, D6- and NS5-branes in the presence of fractional D4-branes. 
This solution is a generalisation of the solution in \cite{Faedo:2020nol} to include fractional D4-branes. In particular, the result is an amalgamation of the background in \cite{Faedo:2020nol} with the technique for including fractional branes used in \cite{Cvetic:2000mh}. 
Before proceeding with constructing the metric let us study the preservation of supersymmetry of such a brane setup. Let $\epsilon_{L,R}$ denote the spinors of type IIA supergravity. They satisfy 
\be
\Gamma^{(10)}\epsilon_{L}=-\epsilon_{L}\, ,\qquad \Gamma^{(10)}\epsilon_{R}=\epsilon_{R}\, ,
\ee
with $\Gamma^{(10)}$ the chirality matrix in 10d. 
For a Dp-brane lying along $012...p$ the supersymmetry is preserved provided the spinors $\epsilon_{L,R}$ satisfy
\be
\epsilon_{L}=\pm \Gamma_{012...p} \epsilon_R\, ,
\ee
with the plus sign for a Dp-brane and the minus for an anti-Dp-brane. For an NS5 brane along $012345$ the projection condition is
\be
\epsilon_L=\Gamma_{012345}\epsilon_L\, ,\qquad \epsilon_R=-\Gamma_{012345}\epsilon_R\, .
\ee
Let us study the brane configuration in table \ref{Table:Dbrane} without specifying whether the branes are anti branes or not by introducing the parameters $\alpha_{\bullet}=\pm 1 $. For the brane setup to preserve supersymmetry the Killing spinors must satisfy
\be
\epsilon_L=\alpha_2 \Gamma_{015}\epsilon_R\, ,\quad \epsilon_{L}=\alpha_2\alpha_6 \Gamma_{6789}\epsilon_L\, ,\quad \epsilon_{L}=\alpha_2 \alpha_6 \Gamma_{01} \epsilon_L\, ,\quad \alpha_2=-\alpha_4\, .
\ee
The last condition implies that for a D2-brane we must include an anti-D4-brane in order to preserve supersymmetry and vice versa. We see that for $\alpha_2\alpha_6=1$ the system preserves $\mathcal{N}=(0,4)$ supersymmetry while for $\alpha_2\alpha_6=-1$ the system preserves $\mathcal{N}=(4,0)$ instead. Without loss of generality we will focus on the $\alpha_2=\alpha_6=1$ case, preserving $\mathcal{N}=(0,4)$ supersymmetry. The projection conditions allow for fractional D4- branes (D6-branes on a shrinking two-cycle) to be included without breaking any further supersymmetry if they wrap certain cycles in the geometry. It follows that the fractional D4-branes (denoted D6') must wrap $01234$ and a two-cycle inside the CY$_2$. This two-cycle must be Poincar\'e dual to a primitive, anti-self-dual $(1,1)$-form in order to preserve the same amount of supersymmetry as the underlying geometry.
We conclude that our setup consists of intersecting NS5-, D2- and D6-branes and anti-D4-branes coupled to fractional D4-branes and preserves $\mathcal{N}=(0,4)$ supersymmetry.

The metric in string frame, following from the brane configuration given in table \ref{Table:Dbrane} and obtained by using the usual supposition rules for intersecting branes, is
\begin{align}
\dd s^2&=\frac{1}{\sqrt{H_{D2}h_{4}H_{D6}h_{8}}}\dd s^2(\mathbb{R}^{1,1})+\frac{\sqrt{H_{D2}H_{D6}}H_{NS5}}{\sqrt{h_{4}h_{8}}}\dd s^{2}(\mathbb{R}^{3})\nonumber\\
&+ \frac{\sqrt{h_{4}h_{8}}H_{NS5}}{\sqrt{H_{D2}H_{D6}}}\dd z^2+ \frac{\sqrt{H_{D2}h_{4}}}{\sqrt{H_{D6}h_{8}}}\dd s^{2}(\mathrm{CY}_{2})\, .
\end{align}
Here the functions $H_{\bullet}$ are harmonic functions on $\mathbb{R}^3$, whilst the function $h_{4}$ depends on both the line interval parametrised by the coordinate $z$ and the Calabi--Yau. In massless Type IIA $h_8$ is a constant while in the massive theory it is promoted to a linear function on the interval, with the leading order piece proportional to the Romans mass. The Bianchi identities for $F_2$ and $F_4$ imply that the functions $H_{\bullet}$ must be equated:
\be
H_{D2}=H_{D6}=H_{NS5}= 1+ \frac{Q}{r}\equiv H(r)\, ,\label{Hequivalence}
\ee
with $r$ the radial distance on $\mathbb{R}^3$.
The metric is supported by the RR fluxes
\begin{align}
F_2&= h_{8} r^2 H'(r)  \dd \vol(S^2)+H_2\, ,\\
F_{4}&=-h_{8} H'(r) \dd \vol(\mathbb{R}^{1,1})\wedge \dd r \wedge \dd z- \partial_{z} h_{4} \dd \vol(CY_2)- h_{8}(\star_{4} \dd_4 h_{4})\wedge \dd z\, ,\nonumber
\end{align}
and by a non-trivial dilaton and NS-NS three-form 
\begin{align}
\me^{-\phi}&=h_8^{5/4} h_{4}^{1/4}\, ,\nonumber\\
\dd B&=r^2 H'(r) \dd z \wedge \dd \vol (S^2)- h_8^{-2} \dd z \wedge H_2\, .
\end{align}
The equations of motion are satisfied provided the function $h_{4}$ satisfies the differential equation\footnote{Note that since $h_8$ is constant here we have redefined the two-form $H_2$, this accounts for the difference with the results in \cite{Lozano:2019emq} which will become manifest when we take the near-horizon limit in the next subsection.}
\be
\partial_{z}^2 h_{4} + h_{8} \nabla^2_{CY_2} h_{4}+ \frac{1}{h_{8}} |H_2|^2=0\,,\label{h4EOM}
\ee
and the two-form $H_2$ is both closed and anti-self-dual living exclusively on the Calabi--Yau two-fold with support along the divisor $C$. The presence of this two-form accounts for the inclusion of the fractional D4 branes. Observe that the condition on $h_4$ is of the form Laplacian plus source term arising from the norm of a form, this is indicative of fractional branes. It is trivial, using the reader's favourite computer program, that this satisfies the equations of motion of massless Type IIA and is supersymmetric.

After equating the $H_{\bullet}$ functions it follows that the metric takes the form
\be
\dd s^2=\frac{1}{\sqrt{h_4 h_8}}\bigg(\frac{1}{H(r)}\dd s^2(\mathbb{R}^{1,1})+ H(r)^2 \dd s^2(\mathbb{R}^3)\bigg)+\frac{\sqrt{h_4}}{\sqrt{ h_8}}\Big(h_8 \dd z^2+ \dd s^2(\text{CY}_{2})\Big)\, .
\ee
The first bracketed part of the metric is that of an extremal five-dimensional black string which has a centre at the poles of the harmonic function $H(r)$. The black string arises from the intersection of the D2- D6- and NS5-branes and not from the D4 branes per se. Instead, the presence of the D4-branes warps the size of the black string through the function $h_4$. Indeed, later we will see that when we take $h_4$ to depend only on the interval we have multiple stacks of D2- and D6-branes stretched between NS5 branes. At the intersection of all three of these types of branes we find a black string. This gives rise to a chain of black strings stacked on top of each other along the interval. It is important to emphasise that this is not a multi-centered black string in $\mathbb{R}^3$ in the usual sense, since such an object has a harmonic function, $H$, which has multiple poles in the radial coordinate. This is not the case here since we take $H$ to have a single pole at $r=0$. We will come back to this point later having constructed the Hanany--Witten brane setup which will make this point manifest. Before proceeding, note that there was nothing special about taking the function $h_4$ to depend only on the interval. In fact more general solutions with Calabi--Yau dependence are possible and are the subject of future work.


\subsection{Near-horizon solution}

Having studied the full solution let us flow to the near-horizon. The near-horizon limit is taken by sending the radial coordinate to zero. After a few trivial rescalings of the coordinates the resulting NS--NS sector of the near-horizon solution is
\begin{align}
\dd s^2&=\frac{1}{\sqrt{h_4 h_8}}\Big( \dd s^2 (\mathrm{AdS}_3) +\frac{1}{4} \dd s^2 (S^2) \Big)+ \sqrt{\frac{h_4}{h_8}} \dd s^2 (\text{CY}_2)+ \sqrt{h_4 h_8} \dd z^2\, , \label{metricNH} \\
\me^{-\Phi}&=  h_4^{1/4} h_8^{5/4}\, ,\quad B=  -\frac{z}{2}\dd \vol(S^2) - \frac{z}{h_8} H_2\, .\label{BNH}
\end{align}
This is supported by the RR fluxes,
\begin{align}
F_2&=-\frac{h_8}{2} \dd \vol(S^2)+ H_2\, ,\label{eq:F2}\\
F_4&= 2 h_{8} \dd z \wedge \dd \vol(\text{AdS}_3)- h_8 \star_4 \dd h_4 \wedge \dd z - \partial_{z} h_4 \dd \vol(\text{CY}_2)\, .\label{eq:F4}
\end{align}
This geometry falls within the classification of $\mathcal{N}=(0,4)$ AdS$_3$ solutions in (massive) Type IIA performed in \cite{Lozano:2019emq}\footnote{More specifically, the  sub-class of $\mathcal{N}=(0,4)$ AdS$_3$ solutions with $u'=0$ is reproduced. See \cite{Dibitetto:2020bsh} for brane solutions leading to AdS$_3$ near horizons with $u'\neq 0$.}. 

A fully explicit solution is provided once a solution to \eqref{h4EOM} is given. In general this is a non-trivial PDE one must solve. A simplifying assumption one may make, is that the function $h_4$ is independent of the Calabi--Yau coordinates. In turn this requires that the norm of the two-form $H_2$ to be constant. It is solutions of this form that we study in this paper and in particular the following section. A more general solution with Calabi--Yau dependence is the subject of future work.


\section{Black string chains} \label{blackstringsglobal}

In this section we will study an explicit class of solutions. We use an ansatz where the function $h_4$ is only a function of the interval and does not depend on the Calabi--Yau coordinates. As explained above, solving equation \eqref{h4EOM} implies that the norm of the two-form $H_2$ on the Calabi--Yau must be constant. We therefore take 
\be
H_2= \gamma \omega\,,\label{H2def}
\ee
with $\omega$ a closed, anti-self-dual two-form on the Calabi--Yau two-fold satisfying
\be
|\omega|^2=1\, ,\quad \int_{C} \omega= 2 \pi,
\ee
with $\omega$ Poincar\'e dual to the divisor $C$ and $\gamma$ taken without loss of generality to be a positive constant.\footnote{One typically normalises a Poincar\'e dual pair such that the integral over the curve of the form is unity. Our normalisation will require additional factors of $\pi$ to appear, in particular the pair $(\omega,C)$ is such that
\begin{equation*}
\int_{C} \alpha= \frac{1}{2\pi} \int_{\text{CY}_2} \alpha \wedge \omega\, ,
\end{equation*}
for any closed two-form $\alpha$ on the Calabi--Yau two-fold.} 
Note that the choice of this two-form uniquely determines the volume of the Calabi--Yau two-fold to be (the slightly uncanonical) $-(2\pi)^2$. One could extend this to include additional divisors, however we will content ourselves with a single divisor in the following. With these assumptions the defining equation reduces to
\be
\partial_{z}^{2} h_4+ \frac{\gamma^{2}}{h_8}=0\, .\label{h4eq}
\ee
This is now a simple linear ordinary differential equation and has general local solution
\be
h_4=\alpha+ \beta z - \frac{\gamma^2}{2h_8}z^2\, ,
\ee
with $\alpha$ and $\beta$ constants. Our goal is to extend this local solution to a global one. This imposes a few constraints on the form of $h_4$. Firstly, in order for the metric to be well defined $h_4$ must be strictly positive, except at the ends of the interval, and moreover it must be continuous. However before we proceed with extending $h_4$ to a global function we must study the Kalb--Ramond two-form $B$ in more detail.


\subsection{Large gauge transformations, NS5 branes and Page fluxes}

For the solution to admit a well-defined partition function in string theory, and not be merely a supergravity solution, the $B$ field must be properly quantised. The condition arises from a generalisation of the Aharonov--Bohm effect for a two-dimensional gauge potential. Recall that the Kalb--Ramond two-form couples minimally to a string with world-sheet $\Sigma$, via 
\be
\frac{1}{(2\pi \ls)^2}\int_{\Sigma} B \, .\label{Aharonov}
\ee
Now a large gauge transformation of the $B$-field by an integral two-form does not change the stringy Aharonov--Bohm phase, since it simply adds $2\pi$ to the phase. Similarly, increasing the flux through $\Sigma$ by multiples of  $8\pi^3 \ls^2$ shifts the stringy Aharonov--Bohm phase by $2\pi$ and is thus the same effect as a large gauge transformation. We are therefore left to conclude that the physically distinct fluxes are those for which \eqref{Aharonov} lies in the interval $[0,1]$. As soon as this condition is violated, we should perform a large gauge transformation by an integral cohomology class.\footnote{The definition includes a normalisation factor of $(2\pi \ls)^{-2}$.} The importance of correctly identifying the large gauge transformation is two-fold. Firstly as we will see it requires the interval to be split into segments, which, upon crossing, NS5 branes are produced, generating a Hanany--Witten like effect. Secondly, the conserved charges in our setup are not Maxwell charges but Page charges which explicitly depend on the choice of gauge for $B$. 
A non-trivial large gauge transformation contributes to the Page charges. Here we will perform this analysis explicitly for our local solution above.

First, let us identify the possible large gauge transformations. Using the representative of the $B$-field given in \eqref{BNH} the necessary large gauge transformations are
\be
B \rightarrow B + \delta B\, ,\qquad \delta B = n\pi \dd \vol(S^2)+ 2 \pi M \omega\, , \label{GenLarge}
\ee
with $n,M$ two constants which parametrise the large gauge transformations and will be fixed later in this section by flux quantisation.
Consider first the two-cycle given by the two-sphere: we have\footnote{We will set the string length, $\ls$ to 1 from now on, however one can reintroduce it by dimensional analysis.}
\be
-\frac{1}{2(2 \pi)^2}\int_{S^2}\Big(2n \pi-z\Big)\dd \vol(S^2)\in [0,1]\, .
\ee
The quantisation condition we must impose is
\be
n - \frac{z}{2\pi} \in [0,1]\, ,
\ee
which implies that we must partition the interval into segments of length $2\pi$. Let the interval have $P+1$ segments of length $2\pi$. As we go from one segment to the next (for increasing $z$) we must perform a large gauge transformation $\delta B= \pi \dd \vol(S^2)$. By a coordinate shift we may take the interval to begin at $z=0$ and consequently the line interval is broken up into the segments
\be
2 \pi k \leq z\leq 2\pi (k+1)\, ,
\ee
for integer $k$. In the $[k,k+1]$ segment the total large gauge transformation we must perform to the representative in equation \eqref{BNH} is 
\be
\delta_{[k]}B = k \pi \dd \vol(S^2)\, .
\ee

Consider now the non-trivial two-cycle $C$ inside the Calabi--Yau two-fold. We have
\be
-\frac{1}{(2\pi)^2} \int_{C}B = - \Big(M- \frac{\gamma}{h_8} \frac{z}{2\pi}\Big)\in [0,1]\, .
\ee
In the interval $[k,k+1]$ we must impose
\be
\frac{\gamma}{h_8} k -M \in [0,1] \quad \text{and} \quad \frac{\gamma}{h_8} (k+1)-M\in [0,1]\, .
\ee
For both to be true it follows that a necessary requirement is $\gamma h_{8}^{-1} \in(0,1]$.\footnote{Note that $h_8$ is necessarily positive in order for the metric to be of correct signature. Despite such a constraint not existing for $\gamma$ we may without loss of generality take it to be positive by a judicious choice of orientation of the two-cycle. } We will see shortly that the ratio must be integer and therefore we must take 
\be
\gamma=h_8\, .
\ee
This however is not the most general way to solve this, an alternative solution is to further partition the line interval, which allows for the solution 
\begin{equation} \label{gamma=ph8}
\gamma=p\, h_8\, ,
\end{equation}
with $p $ a positive integer. The line interval is then divided up into segments of length $2\pi p^{-1}$ 
\be
2 \pi \Big( k+ \frac{l}{p}\Big) \leq z \leq 2 \pi \Big( k+ \frac{l+1}{p}\Big)\, , \quad \l \in \mathbb{Z}\, ,
\ee
with a large gauge transformation proportional to $\omega$ with parameter 
\be
M= k+\frac{l}{p}\, ,
\ee
performed in the $[k,l]$ interval above.

To write the large gauge transformations more concisely it is useful to define some additional notation. Let $\Theta(z-a)$ denote the Heaviside step function defined via
\be
\Theta(z-a)=\begin{cases}
0 & z<a\\
1 & z\geq a
\end{cases}\, ,
\ee
and let, for $a<b$, $\hat{\Theta}$ be given by
\be\label{thetahat}
\hat{\Theta}[a,b]= \Theta(z-2\pi a)-\Theta(z-2\pi b)=\begin{cases} 1 & 2\pi a\leq z\leq 2\pi b\\
0& \text{otherwise}
\end{cases}\, .
\ee
Then, for $\gamma=h_8$, the total large gauge transformation is
\begin{align}
\delta B= \sum_{k=1}^{P} \hat{\Theta}[k,k+1]k(\pi \dd \vol(S^2) +2 \pi  \omega)\, \label{LGBp=1tot}.
\end{align}

Before proceeding with the computation of the Page fluxes let us give the quantisation of the Kalb-Ramond field strength. The field strength is quantised as
\be
\frac{1}{4\pi^2 }\int_{\Sigma_3} H\in \mathbb{Z}\, .
\ee
There are two three-cycles over which we must quantise the flux, both take the form of the product of the interval with one of either the two-sphere or the two-cycle $C$ in the Calabi--Yau Poincar\'e dual to $\omega$. Though the Calabi--Yau admits other two-cycles, the field strength has no support on these cycles and so we may safely ignore them. 

For the three-cycle containing the two-sphere, we find
\begin{align}
\frac{1}{4 \pi^2 } \int_{ I \times S^2} H=P+1\equiv Q_{\text{NS}5}\, .
\end{align}
If we further restrict to each of the segments of length $2\pi$ we see that there is a single NS5 brane in a given segment of the interval. This makes clear the need for the large gauge transformations needed between segments: an NS5 brane is localised on the boundaries of these segments generating a Hanany--Witten like effect. 
Note that this is independent of the choice of $p$, and the further partitioning of the line interval. The large gauge transformation should only be performed when crossing an integer multiple of $2\pi$. 

Now consider the other three-cycle, we find
\be
Q_{\text{NS}5'}\equiv \frac{1}{4 \pi^2 } \int_{I \times C} H=\frac{p}{2 \pi } \int_{0}^{2\pi (P+1)} \dd z= p\, Q_{\text{NS}5}\, .\label{NS5'branes}
\ee
As in the previous case we can look at the number of these NS5' branes in each segment of the interval. It is not difficult to see that in each segment of smallest length (i.e. with segment length $2\pi p^{-1}$) there is precisely 1 NS5' brane. This then accounts for the factor of $p$ in relating the total number of NS5 branes to the NS5' branes in \eqref{NS5'branes}. 

The full analysis of the solution for the $p=1$ case and the $p>1$ case are qualitatively different and we shall only consider the simpler $p=1$ case in this paper.

\vspace{4mm}
\noindent {\bf Page fluxes}

We may now turn our attention to evaluating the Page fluxes of the solution, using the fluxes given in \eqref{eq:F2} and \eqref{eq:F4}. Page fluxes are defined as
\be
\hat{f}\equiv F \wedge \me^{-B}\, ,
\ee
with $F$ denoting the polyform of the \emph{magnetic} parts of the RR-fluxes. Though we require the function $h_4$ to be continuous, so that both the metric and dilaton are well defined, it need not be smooth and may have discontinuities in the first derivative (and higher). Mathematically what we require is that the function is of differentiability class $C^{0}$.\footnote{As we will see shortly the discontinuities give rise to sources for branes and have interesting physics.} We may write the function as the union of a set of smooth continuous functions which have domain each of the segments, subject to a matching condition between segments ensuring that the function is continuous. We take
\be
h_4(z)=\sum_{k} h_{4}^{[k]}(z) \hat{\Theta}[k,k+1]\, ,\label{eq:genh4}
\ee
with $h_{4}^{[k]}$ a smooth continuous function in the interval $2\pi k < z < 2\pi (k+1)$. The continuity of the function implies the matching condition
\be
h_{4}^{[k]}(2\pi k)=h_{4}^{[k-1]}(2\pi k)\, ,
\ee
for each segment.
Explicit computation of the magnetic part of the Page fluxes gives
\begin{align}
\hat{f}_{2}&=\gamma\omega - \frac{h_8}{2} \dd \vol(S^2)\, ,\nonumber\\
\hat{f}_{4}&=-\sum_{k=0}^{P}\Big( \gamma(z-2 \pi k) + h_{4}^{[k]'}\Big) \hat{\Theta}[k,k+1]\dd \vol(\text{CY}_{2})\, , \label{PageFluxp=1}\\
\hat{f}_{6}&=\sum_{k=0}^{P} \frac{1}{2}\Big( h_{4}^{[k]}-(z-2\pi k) h_{4}^{[k]'}- \frac{\gamma}{2}(z-2 \pi k)^2\Big)\hat{\Theta}[k,k+1] \dd \vol(S^2) \wedge \dd \vol(\text{CY}_{2})\, .\nonumber
\end{align}
With these expressions we may compute the Bianchi identities for these Page fluxes, taking care with the step functions. After using the defining equation for $h_4$, namely equation \eqref{h4eq}, we find
\begin{align}
\dd \hat{f}_2&=0\, ,\nonumber\\
\dd \hat{f}_{4}&=\Bigg(-h_{4}^{[0]'}(0)\delta(z)+\sum_{k=1}^{P}\Big(2 \pi \gamma + h_{4}^{[k-1]'}(2\pi k)-h_{4}^{[k]'}(2\pi k)\Big)\delta(z-2\pi k) \nonumber\\
&+\Big(2\pi \gamma+h_{4}^{[P]'}\big(2\pi (P+1)\big)\Big)\delta\big(z-2\pi(P+1)\big)\Bigg)\dd z \wedge \dd \vol(\text{CY}_2)\, ,\label{Bianchip=1}\\\
\dd \hat{f}_{6}&=\sum_{k=1}^{P+1}\pi\Big( h_{4}^{[k-1]'}(2\pi k)+  \pi \gamma\Big)\delta(z-2\pi k)\dd z \wedge \dd \vol(S^2)\wedge \dd \vol(\text{CY})\, .\nonumber
\end{align}
We have simplified the result using the matching conditions and that the function $h_4$ should vanish at the end-points of the line segment. This is a slightly subtle point and we will come back to why this condition is necessary, and also meaningful, later. It is clear that the non-trivial Bianchi identity for $\hat{f}_{4}$ has two different origins. The first comes from the terms with an explicit $\gamma$ factor. This contribution is \emph{new} and purely down to the fractional branes.
The second contribution is a universal contribution and arises when the derivative of $h_4$ is not smooth. The form of this contribution is present even in the absence of fractional branes, however since the function $h_4$ differs in the two cases this contribution also implicitly depends on the fractional branes. For the six-form the origin of the contributions is not as clear cut and we shall postpone this discussion to later in this section.\footnote{Note that in \cite{Lozano:2019zvg} when computing the Bianchi identities the discontinuities in the contributions from the large gauge transformations have been neglected. Taking these into account gives rise to a non-trivial contribution to $\hat{f}_6$. }

\vspace{4mm}
\noindent {\bf Central Charge}

For the final part of our general analysis we will give the formula for the central charge. The Brown--Henneaux formula \cite{Brown:1986nw} specified to our setup reads
\be
c= \frac{3}{2^4 \pi^6 \ls^8}\int \me^{-2 \phi}(h_4 h_8)^{-1/4} \dd \vol_{7}\, .
\ee
For the case at hand where $h_4$ is a function of only the line interval and $h_8$ is constant this may be simplified to
\be
c= \frac{6 h_8}{(2 \pi)^3 \ls^8} \int h_4 \dd z\, .
\ee

\vspace{4mm}


\subsection{Global analysis}

We begin by analysing the gravity solution in more detail: computing the Page charges, the central charge and studying the Bianchi identities. With a thorough understanding of the gravity solution we will construct in the next section a two-dimensional quiver field theory which is conjecturally dual to the gravity solution. We will motivate this two-fold. First by giving a stringy origin using the brane configuration inferred from the gravity analysis and identifying the massless strings stretching between the branes, and secondly by showing that the central charge agrees in the holographic limit with the gravity result.\footnote{We note that the quivers proposed here differ with the ones in \cite{Lozano:2019zvg}, which were conjectured to be the dual of the gravity solutions without fractional branes if we take the $\gamma\rightarrow 0$ limit. We believe that our proposal here is the correct one for both cases and we will explain this using our stringy analysis and by showing that the unnatural scaling argument used in \cite{Lozano:2019zvg} to find a match between both sides of the duality is not needed for our quiver. We will comment more on this in appendix \ref{app:Massive} where we present the corrected quiver for the theory studied in \cite{Lozano:2019zvg}.} 

\vspace{4mm}
\noindent {\bf Gravity Analysis}

To begin we must specify the function $h_4$ explicitly. We have discussed above the continuity constraint however it is necessary to impose some additional constraints on its form in order that the internal space is compact giving a well-defined holographic dual. This is equivalent to imposing that the function $h_4$ has two single roots at both ends of the interval. The reader may be wary of $h_4$ vanishing at the end-points of the interval since it appears prominently in both the metric and dilaton and looks like it will make the metric singular. Indeed, points where $h_4$ vanishes do produce a singularity, however by requiring that $h_4$ has a single root at these points, rather than a double root, one may interpret this singularity as the presence of branes ending the space. Taking the function $h_4$ to have a single root at $z^{*}$ the metric and dilaton close to the root are given by
\begin{align}
\dd s^2 &= \frac{1}{\sqrt{h_8}\sqrt{(z-z^*)h_4'(z^*)}}\Big(\dd s^2 (\text{AdS}_3)+\frac{1}{4}\dd s^2(S^2)\Big)+\frac{\sqrt{(z-z^*)h_4'(z^*)}}{\sqrt{h_8}}\Big( \dd s^2 (\text{CY}_2)+h_8 \dd z^2\Big)\, ,\nonumber\\
\me^{-\Phi}&=\Big((z-z^*)h_{4}'(z^*)\Big)^{1/4}h_8^{5/4}\, ,
\end{align}
which is precisely the form of either an O4 plane or a stack of D4 branes on AdS$_3\times S^2$ smeared over the Calabi--Yau two-fold. For this setup it is only physically sensible to have anti-D4 branes and not O4-planes since smeared O-planes are not well-defined objects in string theory.\footnote{One may wonder whether by allowing dependence of the functions on the CY$_2$ coordinates one can introduce fully localised O4-planes into the geometry and bypass this problem. This will allow for a richer set of solutions and dual field theories and is an interesting problem to consider in the future.}
This is consistent with the field theory description we propose in section \ref{sec:fieldtheory}.
Such an interpretation of the singularity in terms of branes is not possible if $h_4$ has a double root and is therefore not physically interesting. 

\begin{figure}[h]
\begin{center}
\includegraphics{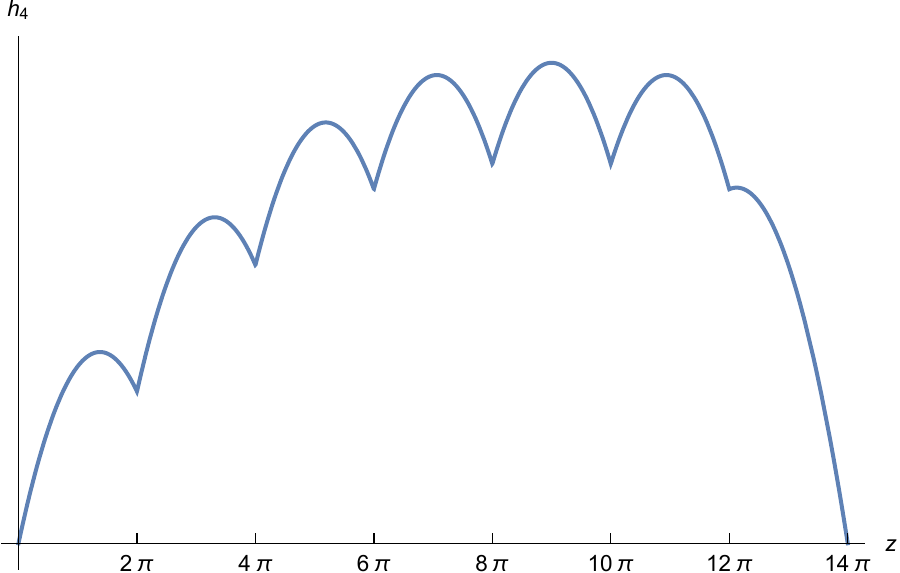}
\caption
{A representative example of $h_4$. Here we have taken $P=6$, and fixed the 7 free parameters as: $\beta_{0}=22\, ,\quad \beta_{1}=21\, ,\quad \beta_{2}=19\, ,\quad \beta_{3}=17\, ,\quad \beta_{4}=15\,,\quad  \beta_{5}=15\, ,\quad\gamma=32$. Note that this example has $\beta_{k-1}\geq \beta_{k}$ for all $k$. }
\label{h4exampleplot}
\end{center}\end{figure}

We may use the shift symmetry in the $z$ coordinate to fix the interval to take values in $[0,2\pi (P+1)]$, with $h_4$ developing a root at both end-points and being strictly positive everywhere else in the domain. An explicit form for $h_4$ following the general expression in \eqref{eq:genh4} is 
\begin{align}\label{h4p=1}
h_4=\begin{cases}
2\pi \beta_{0} z - \frac{\gamma}{2} z^2& 0 \leq z \leq 2\pi\\
(2 \pi)^2 \alpha_{k}+2\pi \beta_{k}(z-2 \pi k)- \frac{\gamma}{2} (z-2 \pi k)^2& 2\pi k \leq z \leq 2 \pi (k+1)\\
(2 \pi)^2 \alpha_{P}+ 2\pi \beta_{P}(z-2 \pi P)- \frac{\gamma}{2} (z-2 \pi P)^2& 2 \pi P \leq z \leq 2\pi (P+1)
\end{cases}\, .
\end{align}
By construction this has a root at $z=0$, however we must still impose that it is both continuous and has a root at $2\pi (P+1)$. 
Continuity requires that the constants are fixed by the iterative constraint,
\be 
\alpha_{k} + \beta_{k} -\frac{\gamma}{2}= \alpha_{k+1}\, ,
\ee
which may be rewritten as
\be \label{alphas}
\alpha_{k}= \sum_{i=0}^{k-1} \Big(\beta_{i}- \frac{\gamma}{2}\Big)\, .
\ee
We see that the fractional branes shift the constants $\alpha_k$.
Notice also that a trivial rewriting gives
\be
\beta_{k-1}-\beta_{k}= 2 \alpha_{k}-\alpha_{k-1}-\alpha_{k+1}\, ,
\ee
which should remind the reader of the anomaly cancellation for certain linear quivers.  
In order for there to be a root at $z=2\pi (P+1)$ we must fix the constants so that 
\be 
\alpha_{P}+ \beta_{P}- \frac{\gamma}{2}=0\, ,\label{betaPm=1}
\ee
or, after using \eqref{alphas}, that
\be
 \sum_{i=0}^{P} \Big(\beta_{i}-\frac{\gamma}{2}\Big)=0\, .\label{p=0betaP}
\ee
Notice that of the $2P+1$ free parameters in \eqref{h4p=1} specifying $h_4$, $P+1$ are fixed in terms of $\gamma$ and the remaining $P$ $\beta$'s. As we will see later, these constraints are essential on the field theory side for anomaly cancellation. The final check is to require that $h_4$ is strictly positive in the domain except at the two end-points where it vanishes. A necessary requirement for this is $\alpha_{k}>0$ for all $1\leq k\leq P$. This immediately implies $\gamma<2 \beta_0$, whilst the constraints on the higher order $\beta$'s are less stringent. We see from \eqref{p=0betaP} that there must be at least one $\beta$, certainly the last one, which satisfies $2 \beta<\gamma$. A representative example of $h_4$ is given in figure \ref{h4exampleplot}.
One should contrast this to the linear rank-functions studied in \cite{Lozano:2019zvg}. The fractional branes lead to a substantially different class of rank function. 

\noindent {\bf Page charges}

With our expression for $h_4$ in hand we are now able to compute the Page charges of the solution. Recall that the magnetic part of the Page fluxes $\hat{f}_{8-p}$, should be quantised according to\footnote{Recall we set $\ls=1$. }
\be
Q_{p} =\frac{1}{(2\pi)^{7-p}} \int_{\Sigma_{8-p}}\hat{f}_{8-p}\, ,
\ee
through all integral cycles. Due to our decomposition of $h_4$ it is natural to look at the various Page charges in each of the segments. The Page charges read
\begin{align} \label{Qcharges}
Q_{D2}^{[k,k+1]}= -\alpha_{k}\, ,\quad
Q_{D4}^{[k,k+1]}= \beta_{k} \, ,\quad
Q_{D6}^{[k,k+1]}= \gamma\, ,\quad
Q_{D6'}^{[k,k+1]}= \gamma\, .
\end{align}
This corroborates that the gravity solution arises from the brane configuration given in table \ref{BraneTablep=1} from which the black string solution was constructed in section \ref{blackstrings}. This will form the starting point for the construction of the field theory dual. Note that the quantisation condition imposes that $\alpha_k,\beta_k, \gamma$ are all integers and therefore consistency with \eqref{alphas} implies that $\gamma$ should be an even integer. 
\begin{table}
\begin{center}
\begin{tabular}{|c|c|c|c|c|c|c|c|c|c|c|c|c|}
\hline
Brane&Page charge&Type&\multicolumn{2}{c|}{$\mathbb{R}^{1,1}$}&\multicolumn{3}{c|}{$\mathbb{R}^3$}&$I$& \multicolumn{4}{c|}{CY$_2$}\\
\hline
 & & &\multicolumn{2}{c|}{}&  \multicolumn{3}{c|}{}& & \multicolumn{2}{c|}{$C$}& \multicolumn{2}{c|}{$\bar{C}$}\\
\hline
\hline
D2& $\alpha_{k}$&Colour&$\times$&$\times$&$-$&$-$&$-$&$\times$&$-$&$-$&$-$&$-$\\
\hline
D4& $\beta_{k}$&Flavour& $\times$&$\times$&$\times$&$\times$&$\times$&$-$&$-$&$-$&$-$&$-$\\
\hline
D6&$\gamma$&Colour&$\times$&$\times$&$-$&$-$&$-$&$\times$&$\times$&$\times$&$\times$&$\times$\\
\hline
D6'&$\gamma$&Flavour&$\times$&$\times$&$\times$&$\times$&$\times$&$-$&$\times$&$\times$&$-$&$-$\\
\hline
NS5&1& N.A.& $\times$&$\times$&$-$&$-$&$-$&$-$ &$\times$&$\times$&$\times$&$\times$\\
\hline
NS5'& 1& N.A. & $\times$&$\times$&$-$&$\times$&$\times$&$-$ &$\times$&$\times$&$-$&$-$\\
\hline
\end{tabular}
\caption{Brane configuration for the solution, with the Page charges associated to the branes in the interval $2\pi k<z<2\pi (k+1)$. We have included the Page charge of the respective branes and indicated whether the branes are flavour or colour. The rule of thumb for distinguishing whether a brane is colour or flavour is to study whether the brane wraps the radial coordinate generating the dilatation symmetry of AdS$_3$ or not. Branes wrapping the radial coordinate give rise to global symmetries on the boundary of AdS$_3$ and thus flavour symmetries whilst those not wrapping the radial coordinate give rise to gauge symmetries, i.e. colour groups for the boundary CFT. }
\label{BraneTablep=1}
\end{center}
\end{table}

We may substitute our expression for $h_4$ into the Bianchi identities computed in \eqref{Bianchip=1}. We find\footnote{Note that we find a different result to that in \cite{Lozano:2019emq} and subsequent follow ups. The reason for this mismatch is because we have taken into account the non-trivial large gauge transformations one must perform in going between segments. The effect of these was neglected in the earlier works and our expressions trivially extend to the $\gamma=0$ case.}
\begin{align}\label{Bianchif4}
\dd \hat{f}_{2}&=0\, ,\\
\dd \hat{f}_{4}&=2\pi\bigg(-\beta_{0}\delta(z)+ \sum_{k=1}^{P}(\beta_{k-1}-\beta_{k}) \delta (z-2 \pi k) + \beta_{P}\delta\big(z-2\pi(P+1)\big)\bigg)\dd z \wedge \dd \vol(\text{CY}_{2})\, ,\nonumber\\
\dd \hat{f}_{6}&=2 \pi^2 \bigg(\sum_{k=1}^{P} (\alpha_{k}-\alpha_{k-1})\delta(z-2\pi k)-\alpha_{P}\delta\big(z-2\pi(P+1)\big) \bigg)\dd z \wedge \dd \vol(S^2) \wedge \dd \vol(\text{CY}_{2})\, .\nonumber
\end{align}
A first point to recall is that the contribution of a Dp-brane and that of an anti-Dp-brane to the source term of the Bianchi identity differs by a minus sign. Since parallel anti-Dp-branes and Dp-branes preserve no supersymmetry\footnote{Concretely the projection condition that the supersymmetry parameters satisfy for a Dp-brane are $\epsilon_{L}=\Gamma_{0...p}\epsilon_{R}$, with the condition for an anti-Dp-brane being $\epsilon_{L}=-\Gamma_{0...p}\epsilon_{R}$. Clearly these are incompatible when the branes are parallel.} we must require that we have either type but not both in our solution. This is equivalent to imposing $\beta_{k-1}-\beta_{k}\geq0$ for all intervals. On the field theory side we will interpret the difference as the rank of the flavour group, which must of course be positive definite and therefore this condition is required on the field theory side also. A second point to stress out is that the first and last contributions to $\dd \hat{f}_{4}$ identify the singularities at both ends of the $z$-interval as associated to O4 orientifold fixed planes, given that both give negative contributions (recall the definition of $\beta_P$ from equation \eqref{betaPm=1}).

We have just seen that the Bianchi identities are satisfied up to the source terms. It remains to interpret these sources. The source terms arise in the Bianchi identities due to the presence of localised branes in the solution. One may derive the general form of the Bianchi identities by supplementing the usual supergravity action with the D-brane effective action
\be
S_{\text{eff}}= S_{\text{DBI}}[g, \phi, B]+ S_{\text{CS}}[C_{p}]\, ,
\ee
for each brane in the theory. We will be concerned with the RR-flux potentials in the following and therefore the DBI part of the action will not play a role in this analysis, only the Chern--Simons part will. 
It takes the form
\be
\mu_{p} \int_{W} \tr \Big(\me^{2 \pi \alpha' \mathcal{F}}\Big)\wedge \sqrt{\frac{\hat{A}(4 \pi^2 \alpha' R_T)}{\hat{A}(4 \pi^2 \alpha' R_{N})}} \wedge \bigoplus_{q} C_{q}\bigg|_{p+1}\, ,
\ee
where $\mathcal{F}$ is the gauge invariant field strength on the Dp-branes to which strings couple, 
\be
2 \pi \alpha' \mathcal{F} =B 1_{n\times n}+ 2 \pi \alpha'\hat{F}\, .
\ee
The polynomial $\hat{A}$ is the A-roof genus and takes as argument the curvature of the tangent and normal bundles respectively. Using the formulation of the potentials which are all electric it is simple to calculate the most general Bianchi identities for the magnetic fluxes. We have\footnote{We use the shorthand $\hat{\mu}_{\bullet}=2 \kappa_{10}^2 \mu_{\bullet}$}
\begin{align}
\dd \hat{f}_2&=\hat{\mu}_6 \tr [1_{D6}] \delta W_{D6}\, ,\\
\dd \hat{f}_4&= \hat{\mu}_{4} \tr[1_{D4}]\delta W_{D4}+2\pi\alpha' \hat{\mu}_{6}\tr [\hat{F}_{D6}]\wedge\delta W_{D6}\, ,\\
\dd \hat{f}_{6}&= \hat{\mu}_{2}\tr[1_{D2}]\delta W_{D2}+2 \pi \alpha' \hat{\mu}_{4}\tr[ \hat{F}_{D4}]\wedge \delta W_{D4}\nonumber\\
&+(2\pi \alpha')^2\hat{\mu}_{6}\Big( \frac{1}{2} \tr [ \hat{F}_{D6}\wedge \hat{F}_{D6}]+(2\pi)^2\tr[1_{D6}]( p_1(R_N)-p_1(R_{T}))\Big)\wedge \delta W_{D6}\, .
\end{align}
Here $\delta W_{\bullet}$ denotes the Poincar\'e dual of the world-volume wrapped by the brane and is normalised to give 1 when integrated over the world-volume. 
Note that both the left- and right-hand side are gauge dependent since it is the gauge dependent field strength $\hat{F}_{\bullet}$ that appears and not the gauge independent combination $\mathcal{F}$. We can now compare these general expressions with the expressions obtained by direct calculation. 

First let us identify the Poincar\'e duals for the various branes, focussing on terms containing delta functions along the interval. Since the D2 and D6 branes are not located at definite points on the interval they will not give rise to such delta function terms along the interval.\footnote{If we had performed this in the full brane solution we would have obtained delta function sources placing these branes at the tip of $\mathbb{R}^3$. In taking the near-horizon limit we have washed out this in the computation of the Bianchi identities. Since our goal is to understand the physics of the line interval we shall ignore such contributions in the following.} Instead, the D4 branes will give rise to delta function terms and we can write
\be 
\delta W_{D4}= -\frac{1}{4\pi}\delta(z-2 \pi k) \dd z \wedge \dd \vol(\text{CY}_2)\, .
\ee
The final contribution we need to consider is from the D6' branes with Poincar\'e dual 
\be
\delta W_{D6'}= \frac{1}{2\pi}\delta(z-2\pi k) \dd z \wedge \omega\, .
\ee
However as we will see momentarily there is no gauge field living on the D6' branes. First let us study the Bianchi identity for $\hat{f}_2$. We see that agreement with $\eqref{Bianchif4}$ implies that there is no brane theory living on the D6' branes located at the distinguished points of the interval. This is not surprising since the number of branes ending on the NS5 brane from the left and right is the same. Given this it follows that the Bianchi identity for $\hat{f}_4$ is satisfied if there is an U$(\beta_{k-1}-\beta_{k})$ gauge theory living on the D4 branes at the $k$'th NS5 brane. Finally the Bianchi identity implies that the field strength on the D4 branes at the $k$'th NS5 brane satisfies
\be
\tr[\hat{F}_{D4}^{(k)}]= \beta_{k}-\frac{\gamma}{2}\, .
\ee
It would be interesting to explicitly construct these fields in the future.

\noindent {\bf Central charge}

Finally, we may compute the central charge of the solution. The Brown--Henneaux formula gives the central charge to be
\begin{align}
c&= \frac{3\pi }{2 G_{N}} \vol(\text{CY}) \int_{0}^{2 \pi (P+1)} \gamma h_4\dd z\nonumber\\
&=6 \gamma  \sum_{k=0}^{P}\Big( \alpha_{k}+\frac{1}{2}\beta_{k}-\frac{\gamma}{6}\Big)\, .
\end{align}
Using the continuity condition we may rewrite this into the form 
\begin{equation}
c=6 \gamma  \sum_{k=0}^{P}\Big( \alpha_{k}+\frac{\gamma}{12}\Big)\, .\label{cp=1hol}
\end{equation}
Since this is a small $\mathcal{N}=(0,4)$ theory the central charge should be an integer multiple of 6. Naively this seems to be problematic since $\gamma$ needs not be divisible by $12$ generically, though it is divisible by 2. The resolution to this apparent paradox is that it is the full central charge, including all subleading contributions, that should be divisible by 6 and not the result from the Brown--Henneaux formula. The gravity calculation we have performed computed the exact leading order piece of the central charge and some, but not all, subleading contributions. If one would in addition compute the full subleading contributions from holography one would find a central charge which is integer including the subleading terms. A similar mechanism was seen in \cite{Couzens:2017way} when considering small $\mathcal{N}=(0,4)$ AdS$_3$ solutions in Type IIB. Only once the subleading contributions were taken into account was the central charge an integer multiple of 6. 

The interpretation of the global solutions constructed in this section is that they describe chains of black strings stacked on top of each other along the $z$-interval. These strings carry three quantised charges, associated to the numbers of D2 and D6 branes wrapped at each interval, together with the number of fractional D4 branes. This is supplemented by global charges that denote the number of D4 branes, flavouring the field theory living on the black string. In the next section we will construct 2d $\mathcal{N}=(0,4)$ quiver CFTs dual to these solutions. These extend the quiver CFTs constructed in \cite{Gadde:2015tra} to include fractional branes. A benefit of our study is that the quivers come along with their explicit $\text{AdS}_3\times S^2\times \text{CY}_2$ holographic duals as studied in this section. This is a well-controlled string theory setting where the implications of holography can be studied in detail.

\section{Field theory}\label{sec:fieldtheory}

\begin{figure}[h!]
\centering
\includegraphics[width=15cm]{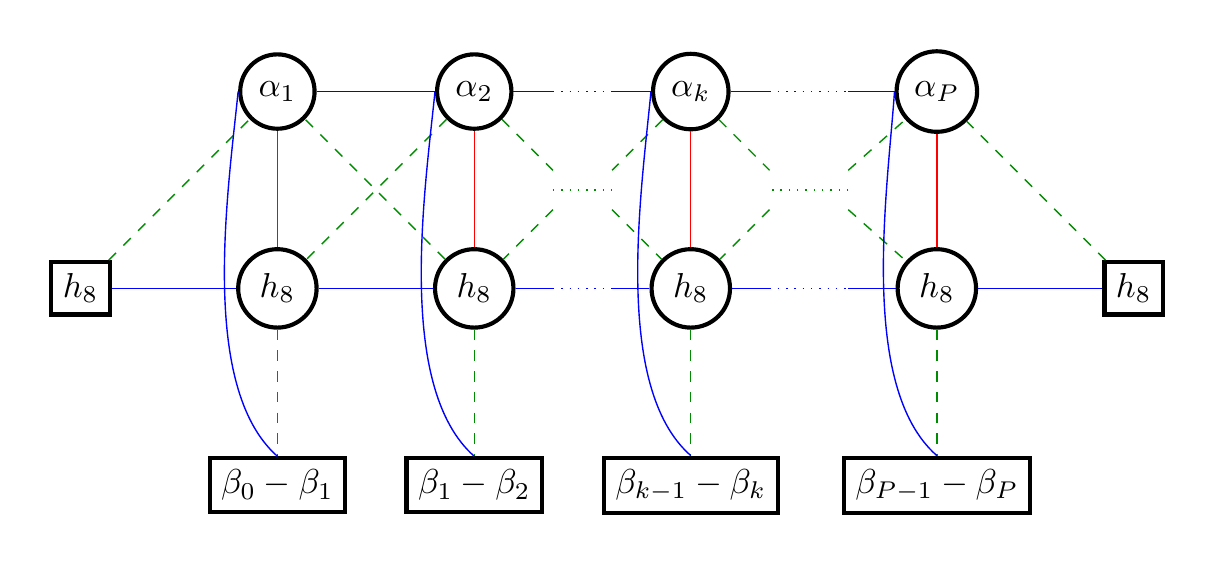}%
\caption{The 2d (0,4) quiver dual to our solutions. The blue lines denote $\mathcal{N}=(4,4)$ twisted hypermultiplets, the red lines denote $\mathcal{N}=(0,4)$ hypermultiplets and green dashed lines are $\mathcal{N}=(0,2)$ Fermi multiplets. The round nodes are $\mathcal{N}=(0,4)$ vector multiplets plus $\mathcal{N}=(0,4)$ adjoint hypermultiplets, whilst the rectangular nodes are flavour symmetries.}
\label{general-quiver-massless}
\end{figure}

Having studied the supergravity solution let us turn our attention to constructing its field theory dual. 
We propose that the field theory dual of the supergravity solutions constructed in the previous section is given by a 2d $\mathcal{N}=(0,4)$ quiver gauge theory with quiver as given in figure \ref{general-quiver-massless}.
The stringy origin of the different multiplets appearing in the quiver is discussed in detail in Appendix \ref{multiplets}.  
Note that our proposal for the quivers has fundamental differences to the quivers previously considered in \cite{Lozano:2019zvg} which did not consider fractional branes. We want to emphasise that the differences are not purely due to the fractional branes, our analysis indicates that the previous quivers are not the correct duals in those cases. To exemplify this we show that the central charge for our new quivers matches the gravity result without the need for unnatural scaling arguments as in the previous works. This analysis is extended to the massive case considered in \cite{Lozano:2019zvg} and later works in Appendix \ref{app:Massive}. The flavour symmetries ending the horizontal line of $h_8$'s in Figure \ref{general-quiver-massless} arise from semi-infinite D6 branes in the geometry. Semi-infinite D6 branes can end on the first and last NS5 branes for a non-zero value of the cosmological constant, created by D8 branes that can be considered to be located far away from the brane system \cite{Hanany:1997sa}. As shown in 
\cite{Cremonesi:2015bld} this is consistent in massless IIA configurations in the large number of nodes limit, ie in the holographic limit.


\subsection{Cancellation of gauge anomalies}

The contribution of the different multiplets in the quiver depicted in Figure \ref{general-quiver-massless}  to the gauge anomaly is given in Table \ref{Table:gaugeanom}. 
\begin{table}[h]
\begin{center}
\begin{tabular}{|c|c|}
\hline
Multiplet & Contribution\\
\hline\hline
(0,4) hyper or twisted hyper (adjoint)& 2N \\
\hline
(0,4) hyper or twisted hyper (fund.) & 1\\
\hline
(0,4) vector  & -2N \\
\hline
(0,2) Fermi& $-\frac12$ \\
\hline
\end{tabular}
\end{center}
\caption{Contribution to the gauge anomaly of the multiplets.}
\label{Table:gaugeanom}
\end{table}

First focus on the $k'$th D2 gauge node. 
It is simple to see that the $\mathcal{N}=(4,4)$ hypermultiplets do not contribute to the anomaly, leaving just the contribution of the $\mathcal{N}=(0,4)$ hypermultiplet and the $(0,2)$ Fermi multiplets. We find that the anomaly condition is trivially satisfied. For the $k$'th D6 gauge node the anomaly is proportional to
\be
2 \alpha_{k}-(\beta_{k-1}-\beta_{k})-\alpha_{k-1}-\alpha_{k+1}\, ,
\ee
which is zero by virtue of the relation
\be
\alpha_{k+1}=\alpha_{k}+\beta_{k}-\frac{\gamma}{2}\, .
\ee
For the end points of the quiver the computation works similarly and is guaranteed by the matching condition. Note that anomaly cancellation is blind to the differences between our quivers and those conjectured in \cite{Lozano:2019zvg} since twisted hypermultiplets and hypermultiplets contribute equally to the anomaly.

\subsection{Central charge}
 
The central charge is given by
\be
c=3 \tr[\gamma^{3} Q^2_{R}]
\ee
with $Q_{R}$ the R-charge under the U$(1)_{R}\subset $ SU(2)$_{R}$, and the trace is over all Weyl fermions in the theory. The matter is organised as in table \ref{R-charges}. 
\begin{table}[h]
\begin{center}
\begin{tabular}{|c|c|c|c|c|}
\hline
Multiplet & $(0,2)$ Origin &  Number of Fermions & Chirality& R-charge of Fermion\\
\hline \hline
$(0,4)$ hyper& 2 $\times$ Chiral& 2 & R.H. & -1\\
\hline
$(0,4)$ twisted hyper& 2$\times$ Chiral & 2 & R.H. & 0\\
\hline
$(0,4)$ vector& (0,2) vector & 1 & L.H. & 1\\
&(0,2) Fermi &1 & L.H. & 1\\
\hline
$(0,2)$ Fermi & -  & 1& L.H. & 0\\
\hline
\end{tabular}
\end{center}
\caption{R-charges and fermion content of the multiplets.}
\label{R-charges}
\end{table}

From the construction of the gauge nodes of our quivers, containing a $\mathcal{N}=(0,4)$ vector multiplet and a $\mathcal{N}=(0,4)$ adjoint hypermultiplet, it is clear that they do not contribute to the R-symmetry anomaly. Moreover, neither the $\mathcal{N}=(0,2)$ Fermi multiplets nor the $\mathcal{N}=(4,4)$ twisted hyper multiplets contribute. Therefore the only contributions to the central charge are from isolated $\mathcal{N}=(0,4)$ hypermultiplets, and the central charge is\footnote{Note that the formula agrees with the standard $\mathcal{N}=(0,4)$ central charge relation $c=6(n_H-n_V)$. Since each $\mathcal{N}=(0,4)$ vector multiplet is accompanied by a $\mathcal{N}= (0,4)$ adjoint hypermultiplet their  contributions to the anomaly cancel. Consequently, the only contributions are from the isolated (by isolated we mean that they do not appear in the completion of a larger multiplet) bifundamental $\mathcal{N}=(0,4)$ hypermultiplets.}
\be
c=3 \cdot 2 n_{H}^{(0,4)}=6 h_{8} \sum_{k=1}^{P}\alpha_{k},\label{eq:CFTc}
\ee
which agrees with the gravity result \eqref{cp=1hol}, upon discarding the $\tfrac{\gamma}{12}$ term which is subleading. Note that since twisted hypermultiplets do not contribute to the anomaly whilst hypermultiplets do one can see the difference between our quiver here and those in \cite{Lozano:2019zvg}. 

Finally we compute the gravitational anomaly. One may compute it using that
\be
c_{L}-c_{R}=\tr \gamma^{3}\, ,\label{Gravanomalyeq}
\ee
where the trace is over the Weyl fermions of the theory as before and $\gamma^3$ is the chirality matrix in 2d. Let us first look at how the multiplets appearing in the quiver contribute. Using table \ref{R-charges} we can see that the gauge nodes contain two right-handed Weyl fermions and two left-handed Weyl fermions and therefore do not contribute to the anomaly. Similarly a $\mathcal{N}=(4,4)$ (twisted) hypermultiplet contains two right-handed Weyl fermions and two left-handed Weyl fermions and therefore also does not contribute to the gravitational anomaly. Conversely, isolated bifundamental $\mathcal{N}=(0,4)$ (twisted-) hypermultiplets contain two right-handed Weyl fermions and therefore contribute $2$ to the anomaly. Finally an isolated $\mathcal{N}=(0,2)$ Fermi multiplet contains a single left-moving Weyl fermion and therefore contributes $-1$ to the anomaly. With these considerations we have the simple formula
\be\label{gravanomaly formula}
c_{L}-c_{R}=2 n_{H}^{(0,4)}- n_{F}^{(0,2)},
\ee
where $n_{\#}^{(p,q)}$ denotes the number of isolated $\mathcal{N}=(p,q)$ $\#$ multiplets in the quiver. We have
\begin{align}
n_{H}^{(0,4)}&= \sum_{k=1}^{P}h_8 \alpha_{k}\, ,\nonumber\\
n_{F}^{(0,2)}&= h_{8}\sum_{k=1}^{P}(\beta_{k-1}-\beta_{k})+2 h_{8}\sum_{k=1}^{P} \alpha_{k}\, ,
\end{align}
and therefore
\be
c_{L}-c_{R}= h_{8}(\beta_{P}-\beta_{0})\, .
\ee
Note that the contribution we get is only from the `end' D4-branes multiplied by the number of D6 branes that they meet. We see that these nodes are somewhat special in that they have only two Fermi lines connecting them whilst the other nodes of this type have three. 

Observe that the gravitational anomaly apparently vanishes for $\beta_{0}=\beta_P$. Recall that we require $\beta_{k-1}\geq \beta_{k}$ and therefore the vanishing of the gravitational anomaly imposes $\beta_{k}=\beta$ for all $k$. Moreover consistency with \eqref{p=0betaP} implies $\beta=\tfrac{\gamma}{2}$, therefore $\alpha_k=0$ and consequently the total central charge, \eqref{eq:CFTc}, vanishes.


\section{Conclusions and future directions}\label{conclusions}

In this paper we have investigated the near-horizon geometry of a chain of 5d black strings living in an asymptotic $\mathbb{R}^{1,4}\times I\times \text{CY}_2$ geometry. The black strings are constructed through a D2-D4-D6-NS5 brane intersection coupled with the presence of fractional branes. We have given a candidate dual 2d quiver CFT, motivated by studying the stringy embedding of the solution, and checked that the central charges of the two sides of the duality are in agreement. One important aspect of this work is that we have provided a different proposal for the dual quiver for the solutions of \cite{Lozano:2019emq} as studied in \cite{Lozano:2019zvg}. Our proposal replaces the need to perform a scaling argument to obtain agreement between the gravity and field theory results. 

There is an interesting alternate construction one can do with the same asymptotic geometry. One may replace the 5d black strings with 5d black holes instead, stacking them along the interval. The near-horizon geometries of this setup, which will contain an AdS$_2$ factor, were constructed in \cite{Lozano:2020txg,Lozano:2020sae,Lozano:2021rmk}. There AdS$_2$ solutions in Type II supergravities were constructed from the seed solutions studied in \cite{Lozano:2019emq} (which is also the seed for the solution in this paper) using both double-analytic continuation and T-duality. Like here, they possess a closed two-form $H_2$ and a function satisfying a similar equation to the defining equation here. Given the origin of these solutions, and the close connection to those of this paper, it would be interesting to study these solutions in this light extending known explicit examples and their dual field theories.

Another interesting setup, connected to the work here, are AdS$_3$ solutions in M-theory with $\mathcal{N}=(0,4)$ supersymmetry which are the uplifts of the solutions discussed here. These solutions should arise from taking the near-horizon of chains of 6d black strings living in the asymptotic geometry $\mathbb{R}^{1,5}\times I\times \text{CY}_{2}$. The most general solutions constructed in \cite{Lozano:2020bxo} contain fractional M5-branes, arising from KK-monopoles wrapping collapsing two-cycles in the Calabi--Yau two-fold, that should modify the quivers constructed in \cite{Lozano:2020bxo} in a manner similar to that of this paper. 

Moreover, one can consider backgrounds in M-theory of the type $\mathbb{R}^{1,3}\times I\times \text{CY}_3$ and look for BPS solutions in which 4d black holes are stacked along the interval. If such solutions exist, they will have a near horizon geometry AdS$_2 \times S^2 \times I \times \text{CY}_3$ that might be related to the solutions recently constructed in \cite{Lozano:2021fkk} (see appendix A). A final direction one may consider are generalising this setup to rotating black strings. The rotation requires the near-horizon geometry to include a non-trivial fibration of the symmetries of the internal manifold over AdS$_3$, thereby evading the classification in \cite{Lozano:2019emq} of AdS$_3$ solutions that we have based this work on. In type IIB a sub-class of possible solutions have been classified in \cite{Couzens:2020jgx}.
We leave these interesting directions for future work.

\section*{Acknowledgments}

We would like to thank Carlos Nunez for very useful discussions. C.C. would like to thank KIAS for hospitality during the closing stages of this work. Y.L. and N.P. are partially supported by the Spanish government grant PGC2018-096894-B-100. C.C. and S.V are supported in part by the D-ITP consortium, a program of the Netherlands Organisation for Scientific Research  (NWO) that is funded by the Dutch Ministry of Education, Culture and Science (OCW).

\appendix

\section{Multiplets from strings}\label{multiplets}

In this appendix we give a stringy origin for the quiver theory depicted in Figure \ref{general-quiver-massless}. Our analysis utilises the brane intersection underlying the gravity solution, from which one can study the different ways of obtaining massless modes from strings stretching between the branes. We view this as giving additional weight to the conjectured field theory dual of our supergravity setup alongside the matching of the central charges. 

Recall that the branes live in the asymptotic geometry
\be
\mathbb{R}^{1,4} \times I \times \text{CY}_2\, ,
\ee
and are configured as given in table \ref{BraneTablep=1 app}.
\begin{table}
\begin{center}
\begin{tabular}{|c|c|c|c|c|c|c|c|c|c|c|c|}
\hline
Brane&Type&\multicolumn{2}{c|}{$\mathbb{R}^{1,1}$}&\multicolumn{3}{c|}{$\mathbb{R}^3$}&$I$& \multicolumn{4}{c|}{CY$_2$}\\
\hline
 & & \multicolumn{2}{c|}{}& $r$ &\multicolumn{2}{c|}{$S^2$}& & \multicolumn{2}{c|}{$C$}& \multicolumn{2}{c|}{$\bar{C}$}\\
\hline
\hline
D2&Colour&$\times$&$\times$&$-$&$-$&$-$&$\times$&$-$&$-$&$-$&$-$\\
\hline
D4&Flavour& $\times$&$\times$&$\times$&$\times$&$\times$&$-$&$-$&$-$&$-$&$-$\\
\hline
D6&Colour&$\times$&$\times$&$-$&$-$&$-$&$\times$&$\times$&$\times$&$\times$&$\times$\\
\hline
D6'&Flavour&$\times$&$\times$&$\times$&$\times$&$\times$&$-$&$\times$&$\times$&$-$&$-$\\
\hline
NS5& & $\times$&$\times$&$-$&$-$&$-$&$-$ &$\times$&$\times$&$\times$&$\times$\\
\hline
NS5'&  & $\times$&$\times$&$-$&$\times$&$\times$&$-$ &$\times$&$\times$&$-$&$-$\\
\hline
\end{tabular}
\caption{Brane configuration for the solution.}
\label{BraneTablep=1 app}
\end{center}
\end{table}
The presence of the branes, and compact Calabi--Yau space, breaks the SO$(1,9)$ Lorentz group to SO$(1,1)\times$SO$(3)$. Recall that the double cover of SO$(3)$, SU$(2)_R$ is dual to the R-symmetry of the SCFT. When the interval is periodically identified the symmetry group is enhanced by a flavour U$(1)$ coming from the now circular interval. Given that the function $h_4$ is quadratic in the interval coordinate it follows that in this case when it is periodically identified it must be constant and the fractional branes disappear. One may then T-dualise along this U$(1)$ to obtain the D1-D5 system. Since this is well-studied we shall ignore this case with enhanced symmetry and assume a non-trivial interval in the remainder of this section.

The key to this analysis is in identifying the 2d multiplets one obtains from quantising the fundamental strings stretching between the various branes in the setup. There is a large literature on this type of analysis for similar setups, much of which is transferable to this setup, see for example \cite{Tong:2014yna,Hanany:1996ie}. Despite this, we will be as detailed as possible in order to present a consistent and complete story at the cost of reviewing some `well-known' material in places. We hope the reader can forgive us for this, but for those who just want to jump to the punch-line we have presented an overall summary in table \ref{table:multiplets from strings}. 

\begin{table}[h]

\begin{center}

\begin{tabular}{|c|c|c|c|}

\hline

String & Segment & Multiplet & Representation\\
\hline\hline

D2-D2 & Same & $\mathcal{N}=(0,4)$ vector $+$ $\mathcal{N}=(0,4)$ hyper & Adjoint\\
\hline
D2-D2 & Adjacent & $\mathcal{N}=(4,4)$ twisted hyper & bi-fundamental\\
\hline
D6-D6 & Same & $\mathcal{N}=(0,4)$ vector $+$ $\mathcal{N}=(0,4)$ hyper & Adjoint\\
\hline
D6-D6 & Adjacent & $\mathcal{N}=(4,4)$ twisted hyper & bi-fundamental\\
\hline
D2-D6 & Same &$\mathcal{N}=(0,4)$ hyper & bi-fundamental\\
\hline
D2-D6 & Adjacent & $\mathcal{N}=(0,2)$ Fermi & bi-fundamental\\
\hline
D2-D4 & Same & $\mathcal{N}=(4,4)$ twisted hyper & bi-fundamental\\
\hline
D4-D6 & Same & $\mathcal{N}=(0,2) $ Fermi & bi-fundamental\\
\hline

\end{tabular}

\end{center}

\caption{We give the summary of the multiplets arising from the different strings stretching from the branes in the setup. The segment column determines whether the branes lie in the same segment or in adjacent segments.
For strings that do not contribute massless modes we have ignored their contribution in the table, for example D4-D4 strings.}

\label{table:multiplets from strings}

\end{table}

\subsection*{D2-D2 strings}

There are two distinct cases of D2-D2 strings to consider depending on whether both end-points of the string lie in the same segment of the line interval or not. We first consider the case where they are in the same segment before considering adjacent segments. For segments which are not adjacent there are no massless modes on the strings and therefore these will be ignored in the following. 

For the case where the two endpoints both lie in the same segment the problem reduces to identifying the massless modes of a stack of D2-branes bounded in one spacetime direction by NS5 branes. This is a well-studied problem, see for example the T-dual setup in \cite{Hanany:1996ie}, however for completeness let us sketch the argument as there is still an important point we wish to emphasise. On a stack of $N$ infinitely extended D2 branes there lives a 3d U$(N)$ gauge theory consisting of a single $\tfrac{1}{2}$-BPS vector multiplet. We now want to bound the D2 brane in one spacetime direction by two NS5 branes, one at each end-point. The D2-branes are now infinite in extent in only two spacetime dimensions and the effective theory living on the branes becomes two-dimensional. Furthermore, the presence of the NS5-branes breaks one half of the supersymmetry of the setup.\footnote{To see this use that the preserved supersymmetry parameters of a D2 brane lying along $012$ satisfy $\epsilon_{L}=\Gamma_{012}\epsilon_R$ whilst for an NS5 brane lying along $013456$ and localised at fixed points in $2$, the preserved supersymmetry parameters satisfy $\epsilon_{L}=\Gamma_{013456}\epsilon_L$ and $\epsilon_{R}=-\Gamma_{013456}\epsilon_{R}$. Here $3456$ span the directions of the Calabi--Yau and $2$ is the direction of the interval.}  

We now want to decompose the 3d vector multiplet in terms of 2d multiplets before truncating out the fields fixed by the Neumann boundary conditions affixing the D2 branes to the NS5 branes. Recall that the bosonic field content of a $\tfrac{1}{2}$-BPS 3d vector multiplet is a 3d vector and 7 real scalars: the latter of which parametrise the fluctuations of the D2 brane in the transverse directions. Reducing to 2d we end up with a 2d vector and 8 real scalars, the eighth scalar coming from the third component of the 3d vector in the KK reduction. The preserved supersymmetry arising from the projection conditions implies that the fields combine into: a $\mathcal{N}=(0,4)$ vector multiplet; a $\mathcal{N}=(0,4)$ hypermultiplet, arising from the fluctuations along the Calabi--Yau; and a $\mathcal{N}=(4,4)$ twisted hypermultiplet arising from combining the scalar in the decomposition of the 3d vector and the fluctuations in $\mathbb{R}^3$.\footnote{To distinguish between the scalars forming hypermultiplets or twisted hypermultiplets one should consider the transformation properties of the fields under the R-symmetry of the solution. In the present setup the R-symmetry is SU$(2)$ which is the double cover of the SO$(3)$ rotations acting on $\mathbb{R}^3$. We see then that the fluctuations along the Calabi--Yau directions should be singlets under the SU$(2)$ and therefore neutral under the R-symmetry and consequently hypermultiplets. Conversely the fluctuations arising from the KK-reduction of the gauge field and the fluctuations in $\mathbb{R}^3$ are charged under the R-symmetry and are therefore twisted hypermultiplets.} In order for the D2 brane to end on the NS5's we must impose Neumann boundary conditions on the fields. This sets to zero the $\mathcal{N}=(4,4)$ twisted hypermultiplet, leaving just the $\mathcal{N}=(0,4)$ vector multiplet and the $\mathcal{N}=(0,4)$ hypermultiplet, in the adjoint of U$(N)$. Note that this does not combine into a $\mathcal{N}=(4,4)$ vector multiplet. This is one difference between our quiver here and the quivers appearing previously in  \cite{Lozano:2019zvg}, where the nodes were taken to be $\mathcal{N}=(4,4)$ vector multiplets. 

Having identified the multiplets from a stack of D2 branes in the same segment let us consider D2 branes in adjacent segments. Strings stretching between D2-branes which are not adjacent are massive, the massless modes are located at the intersection of the two D2-branes with the NS5 brane, but what are these massless modes? From the structure of the Chan-Paton factors of the two end-points it is clear that this must be bi-fundamental matter. The two D2-branes must meet on the NS5 brane, this fixes the degrees of freedom moving in the Calabi--Yau directions. What remains is to move in the directions of $\mathbb{R}^3$ and following \cite{Hanany:1996ie} we obtain scalars transforming in the $2$ of SU($2)_R$ which combine into a twisted-hypermultiplet. This hypermultiplet is in fact a $\mathcal{N}=(4,4)$ twisted hypermultiplet as can be seen by studying the supersymmetry parameters of the brane setup. 

We have seen that the D2-D2 strings furnish our quiver with two types of matter multiplets. We have gauge nodes containing a $\mathcal{N}=(0,4)$ vector multiplet and a $\mathcal{N}=(0,4)$ hypermultiplet in the adjoint, whilst the nodes are connected to the adjacent ones via bi-fundamental $\mathcal{N}=(4,4)$ twisted hypermultiplets.

\subsection*{D6-D6 strings}

Next consider the D6-D6 strings. This is in fact T-dual (or mirror symmetric) to the case of the D2-D2 strings considered previously. Therefore the spectrum is the same. We have gauge nodes with a $\mathcal{N}=(0,4)$ vector multiplet and $\mathcal{N}=(0,4)$ hypermultiplet in the adjoint, connected to adjacent nodes by bi-fundamental $\mathcal{N}=(4,4)$ twisted hypermultiplets.

\subsection*{D2-D4 strings}

For the D2-D4 strings we may again make use of a duality to obtain a well-studied setup. T-dualising along one Calabi--Yau direction one obtains a stack of D3-branes ending on a D5-brane whose world-volume contains the $\mathbb{R}^3$ factor. This is precisely the setup in \cite{Hanany:1996ie} and the massless modes are the position of the D3-branes inside the D5 brane. This gives rise to a bi-fundamental $\mathcal{N}=(4,4)$ twisted hypermultiplet.

\subsection*{D6-D4 strings}

For the D6-D4 strings note that the massless modes arise from the fluctuations of the string end-points stretching between the two branes. Since all bosonic degrees of freedom are fixed in order that the two branes intersect there are no fluctuations and therefore no bosonic zero modes. Instead we obtain a Fermi multiplet. We may see this more concretely by noting that the setup is T-dual to the D0-D8 system. Using \cite{Banks:1997zs} we see that the strings give rise to a bifundamental $\mathcal{N}=(0,2)$ Fermi multiplet.

\subsection*{D2-D6 strings}

The final fundamental strings that we must consider are the D2-D6 strings. We have two cases to consider. When the two branes are in the same interval the setup is T-dual to the D0-D4 where the D4 is wrapped on the Calabi--Yau two-fold. This gives rise to a $\mathcal{N}=(0,4)$ hypermultiplet, parametrising the location of the instanton within the D4 brane world-volume.

Finally, when strings stretch between a D2 and D6 brane in adjacent intervals we may T-dualise to a D1-D5 system where we view the D1 as ending on the D5 which is wrapped on the Calabi--Yau two-fold. This gives rise to a $\mathcal{N}=(0,2)$ Fermi multiplet.

\section{Massive quivers}\label{app:Massive}

In the main text we have emphasised that the quivers we propose to be dual to the geometries studied in section \ref{blackstringsglobal} differ from those studied in similar setups in \cite{Lozano:2019zvg}. In this appendix we will apply the knowledge learnt in studying our setup to correct the massive quivers first studied in that reference. We will set the fractional branes to vanish and reinstate the non-trivial Romans mass which leads to new flavour D8-branes.\footnote{Due to the different completions of the solutions there is no limit in which one can recover the massless case without fractional branes from the massive case studied in this appendix. The issue arises because the quiver in the massive case studied here is the one obtained when there are both D4 and D8 branes ending the space and there is no consistent way of removing these D8 branes in trying to take a massless limit. One may instead consider a massive quiver where the space does not end with D8 branes, that is, the function $h_8$ does not vanish at the end-points in the dual gravity solution. However this modifies the quiver studied in \cite{Lozano:2019zvg} which is the quiver we will discuss here.} As in the previous section, let us study the massless multiplets arising from the strings stretching between the various branes of the setup. We may use the previous analysis and supplement it with the new multiplets arising from strings stretching between the D8-branes to the D2-, D4-, and D6-branes.  

\begin{figure}[h!]
\centering
\includegraphics[width=12cm]{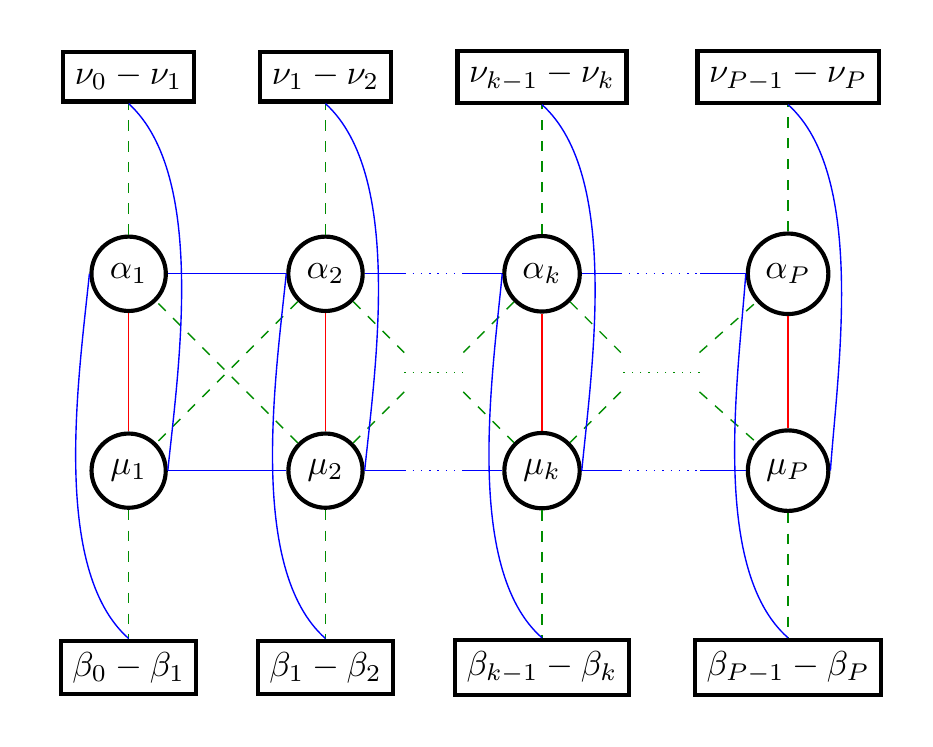}%
\caption{The 2d $\mathcal{N}=(0,4)$ quiver dual to the massive Type IIA solutions. As before, the blue lines denote $\mathcal{N}=(4,4)$ twisted hypermultiplets, the red lines denote $\mathcal{N}=(0,4)$ hypermultiplets and green dashed lines are $\mathcal{N}=(0,2)$ Fermi multiplets. The round nodes are $\mathcal{N}=(0,4)$ vector multiplets plus $\mathcal{N}=(0,4)$ adjoint hypermultiplets, whilst the rectangular nodes are flavour symmetries. }
\label{general-quiver-massive}
\end{figure}

First, the D2-D8 system is T-dual to the D6-D4 system studied above and we conclude that the D2-D8 strings give rise to a bifundamental $\mathcal{N}=(0,2)$ Fermi multiplet. Similarly the D6-D8 system is T-dual to the D2-D4 system studied above and therefore we have a $\mathcal{N}=(4,4)$ twisted hypermultiplet. Finally, strings stretching between the D4 and D8-branes do not contribute massless multiplets to the quiver. Having identified the massless modes we can construct the quiver, which is given in figure \ref{general-quiver-massive}. 

Let us now check that our proposal is firstly anomaly free and secondly that the central charge matches the gravitational result obtained in \cite{Lozano:2019zvg}.
Using similar arguments to the case considered in the main text, and the relations 
\be
\beta_{k}=\alpha_{k+1}-\alpha_{k}\, ,\quad \nu_{k}=\mu_{k+1}-\mu_{k}\, ,\label{massivecontraints}
\ee
it follows that the quiver is anomaly free.

The Brown--Henneaux formula for the central charge gives the simple result
\be
c_{BH}=\sum_{k=0}^{P}\Big(6 \alpha_{k}\mu_{k} +3 (\mu_k \beta_k+\alpha_k \nu_k)+2 \beta_k \nu_k\Big)\, .
\ee
The first thing to note is that only the first term is generically an integer multiple of $6$, despite being dual to a $\mathcal{N}=(0,4)$ SCFT. As we pointed out earlier this is not necessarily a contradiction since Brown--Henneaux does not give $c_{R}$ but the sum \cite{Kraus:2005zm}
\be
c_{BH}=\frac{c_{L}+c_{R}}{2}\, ,
\ee
and furthermore higher derivative corrections may still contribute to the subleading terms. 
Using the constraints \eqref{massivecontraints}
we may write
\be\label{eq:massivecgrav}
c_{BH}=\sum_{k=0}^{P}\Big( 6 \alpha_{k}\mu_{k}+\alpha_{k}\mu_{k+1}+\alpha_{k+1}\mu_{k}-2 \alpha_{k}\mu_{k}\Big)\, .
\ee

We now want to compare this with the field theory result, keeping in mind that this is \emph{not} the $c_{R}$ that one computes in field theory. 
Following our earlier discussion, the contribution to the central charge from the field theory is given by 6 times the number of isolated $\mathcal{N}=(0,4)$ hypermultiplets. For the quiver at hand this is
\be
c_R=6 \sum_{k=1}^{P} \alpha_{k}\mu_{k}\, .\label{cRmassive}
\ee
We can also compute the gravitational anomaly using \eqref{gravanomaly formula}. We find
\begin{align}
c_{L}-c_{R}&=\sum_{k=1}^{P}\Big(\alpha_{k}(\mu_{k+1}-\mu_{k})+\mu_{k}(\alpha_{k+1}-\alpha_{k})\Big)\nonumber\\
&=\sum_{k=1}^{P}\Big(\alpha_{k}\mu_{k+1}+\alpha_{k+1}\mu_{k}-2 \alpha_{k}\mu_{k}\Big)\, .\label{gravmassive}
\end{align}
Finally combining \eqref{cRmassive} and \eqref{gravmassive} we should compare the gravity result to
\begin{align}
c_{BH}^{\text{CFT}}=c_{R}+\frac{1}{2}(c_{L}-c_{R})&= \sum_{k=1}^{P}\Big(6 \alpha_{k}\mu_{k}+\frac{1}{2}(\alpha_{k}\mu_{k+1}+\alpha_{k+1}\mu_{k}-2\alpha_{k} \mu_{k})\Big)\nonumber\\
&=\sum_{k=1}^{P}\Big(6 \alpha_{k}\mu_{k}+ \frac{1}{2}
\big(\alpha_{k}(\mu_{k+1}-\mu_{k})+\mu_{k}(\alpha_{k+1}-\alpha_{k})\big)\Big)\, .
\end{align}
We note that the leading order terms match exactly without the need for any scaling argument as used in the literature previously. We believe that once higher order corrections to the gravity computation are taken into account one will obtain a match even at subleading order. One curiosity of our results is that the central charge as computed from gravity, \eqref{eq:massivecgrav}, is exactly $c_{L}$. We leave understanding this and the higher derivative corrections to future work.

\bibliographystyle{JHEP}
\bibliography{BlackStringsthefinalbib}

\providecommand{\href}[2]{#2}\begingroup\raggedright\begin{thebibliography}{10}

\bibitem{Lozano:2019emq}
Y.~Lozano, N.~T. Macpherson, C.~Nunez and A.~Ramirez, \emph{{AdS$_3$ solutions
  in Massive IIA with small $\mathcal{N}=(4,0)$ supersymmetry}},
  \href{http://dx.doi.org/10.1007/JHEP01(2020)129}{\emph{JHEP} {\bf 01} (2020)
  129}, [\href{https://arxiv.org/abs/1908.09851}{{\tt 1908.09851}}].

\bibitem{Strominger:1996sh}
A.~Strominger and C.~Vafa, \emph{{Microscopic origin of the Bekenstein-Hawking
  entropy}}, \href{http://dx.doi.org/10.1016/0370-2693(96)00345-0}{\emph{Phys.
  Lett. B} {\bf 379} (1996) 99--104},
  [\href{https://arxiv.org/abs/hep-th/9601029}{{\tt hep-th/9601029}}].

\bibitem{Maldacena:1997de}
J.~M. Maldacena, A.~Strominger and E.~Witten, \emph{{Black hole entropy in M
  theory}}, \href{http://dx.doi.org/10.1088/1126-6708/1997/12/002}{\emph{JHEP}
  {\bf 12} (1997) 002}, [\href{https://arxiv.org/abs/hep-th/9711053}{{\tt
  hep-th/9711053}}].

\bibitem{Vafa:1997gr}
C.~Vafa, \emph{{Black holes and Calabi-Yau threefolds}},
  \href{http://dx.doi.org/10.4310/ATMP.1998.v2.n1.a8}{\emph{Adv. Theor. Math.
  Phys.} {\bf 2} (1998) 207--218},
  [\href{https://arxiv.org/abs/hep-th/9711067}{{\tt hep-th/9711067}}].

\bibitem{Minasian:1999qn}
R.~Minasian, G.~W. Moore and D.~Tsimpis, \emph{{Calabi-Yau black holes and
  (0,4) sigma models}}, {\emph{Commun. Math. Phys.} {\bf 209} (2000) 325--352},
  [\href{https://arxiv.org/abs/hep-th/9904217}{{\tt hep-th/9904217}}].

\bibitem{Castro:2008ne}
A.~Castro, J.~L. Davis, P.~Kraus and F.~Larsen, \emph{{String Theory Effects on
  Five-Dimensional Black Hole Physics}},
  \href{http://dx.doi.org/10.1142/S0217751X08039724}{\emph{Int. J. Mod. Phys.
  A} {\bf 23} (2008) 613--691}, [\href{https://arxiv.org/abs/0801.1863}{{\tt
  0801.1863}}].

\bibitem{Haghighat:2013tka}
B.~Haghighat, C.~Kozcaz, G.~Lockhart and C.~Vafa, \emph{{Orbifolds of
  M-strings}}, \href{http://dx.doi.org/10.1103/PhysRevD.89.046003}{\emph{Phys.
  Rev. D} {\bf 89} (2014) 046003}, [\href{https://arxiv.org/abs/1310.1185}{{\tt
  1310.1185}}].

\bibitem{Gadde:2015tra}
A.~Gadde, B.~Haghighat, J.~Kim, S.~Kim, G.~Lockhart and C.~Vafa, \emph{{6d
  String Chains}}, \href{http://dx.doi.org/10.1007/JHEP02(2018)143}{\emph{JHEP}
  {\bf 02} (2018) 143}, [\href{https://arxiv.org/abs/1504.04614}{{\tt
  1504.04614}}].

\bibitem{Haghighat:2015ega}
B.~Haghighat, S.~Murthy, C.~Vafa and S.~Vandoren, \emph{{F-Theory, Spinning
  Black Holes and Multi-string Branches}},
  \href{http://dx.doi.org/10.1007/JHEP01(2016)009}{\emph{JHEP} {\bf 01} (2016)
  009}, [\href{https://arxiv.org/abs/1509.00455}{{\tt 1509.00455}}].

\bibitem{Lawrie:2016axq}
C.~Lawrie, S.~Schafer-Nameki and T.~Weigand, \emph{{Chiral 2d theories from N =
  4 SYM with varying coupling}},
  \href{http://dx.doi.org/10.1007/JHEP04(2017)111}{\emph{JHEP} {\bf 04} (2017)
  111}, [\href{https://arxiv.org/abs/1612.05640}{{\tt 1612.05640}}].

\bibitem{Couzens:2017way}
C.~Couzens, C.~Lawrie, D.~Martelli, S.~Schafer-Nameki and J.-M. Wong,
  \emph{{F-theory and AdS$_{3}$/CFT$_{2}$}},
  \href{http://dx.doi.org/10.1007/JHEP08(2017)043}{\emph{JHEP} {\bf 08} (2017)
  043}, [\href{https://arxiv.org/abs/1705.04679}{{\tt 1705.04679}}].

\bibitem{Couzens:2019wls}
C.~Couzens, H.~het Lam, K.~Mayer and S.~Vandoren, \emph{{Black Holes and (0,4)
  SCFTs from Type IIB on K3}},
  \href{http://dx.doi.org/10.1007/JHEP08(2019)043}{\emph{JHEP} {\bf 08} (2019)
  043}, [\href{https://arxiv.org/abs/1904.05361}{{\tt 1904.05361}}].

\bibitem{Bena:2006qm}
I.~Bena, D.-E. Diaconescu and B.~Florea, \emph{{Black string entropy and
  Fourier-Mukai transform}},
  \href{http://dx.doi.org/10.1088/1126-6708/2007/04/045}{\emph{JHEP} {\bf 04}
  (2007) 045}, [\href{https://arxiv.org/abs/hep-th/0610068}{{\tt
  hep-th/0610068}}].

\bibitem{Couzens:2020aat}
C.~Couzens, H.~het Lam, K.~Mayer and S.~Vandoren, \emph{{Anomalies of (0,4)
  SCFTs from F-theory}},
  \href{http://dx.doi.org/10.1007/JHEP08(2020)060}{\emph{JHEP} {\bf 08} (2020)
  060}, [\href{https://arxiv.org/abs/2006.07380}{{\tt 2006.07380}}].

\bibitem{Haghighat:2013gba}
B.~Haghighat, A.~Iqbal, C.~Koz\c{c}az, G.~Lockhart and C.~Vafa,
  \emph{{M-Strings}},
  \href{http://dx.doi.org/10.1007/s00220-014-2139-1}{\emph{Commun. Math. Phys.}
  {\bf 334} (2015) 779--842}, [\href{https://arxiv.org/abs/1305.6322}{{\tt
  1305.6322}}].

\bibitem{Lockhart:2012vp}
G.~Lockhart and C.~Vafa, \emph{{Superconformal Partition Functions and
  Non-perturbative Topological Strings}},
  \href{http://dx.doi.org/10.1007/JHEP10(2018)051}{\emph{JHEP} {\bf 10} (2018)
  051}, [\href{https://arxiv.org/abs/1210.5909}{{\tt 1210.5909}}].

\bibitem{Kim:2012qf}
H.-C. Kim, J.~Kim and S.~Kim, \emph{{Instantons on the 5-sphere and
  M5-branes}},  \href{https://arxiv.org/abs/1211.0144}{{\tt 1211.0144}}.

\bibitem{Martelli:2003ki}
D.~Martelli and J.~Sparks, \emph{{G structures, fluxes and calibrations in M
  theory}}, \href{http://dx.doi.org/10.1103/PhysRevD.68.085014}{\emph{Phys.
  Rev. D} {\bf 68} (2003) 085014},
  [\href{https://arxiv.org/abs/hep-th/0306225}{{\tt hep-th/0306225}}].

\bibitem{Tsimpis:2005kj}
D.~Tsimpis, \emph{{M-theory on eight-manifolds revisited: N=1 supersymmetry and
  generalized spin(7) structures}},
  \href{http://dx.doi.org/10.1088/1126-6708/2006/04/027}{\emph{JHEP} {\bf 04}
  (2006) 027}, [\href{https://arxiv.org/abs/hep-th/0511047}{{\tt
  hep-th/0511047}}].

\bibitem{Kim:2005ez}
N.~Kim, \emph{{AdS(3) solutions of IIB supergravity from D3-branes}},
  \href{http://dx.doi.org/10.1088/1126-6708/2006/01/094}{\emph{JHEP} {\bf 01}
  (2006) 094}, [\href{https://arxiv.org/abs/hep-th/0511029}{{\tt
  hep-th/0511029}}].

\bibitem{Kim:2007hv}
H.~Kim, K.~K. Kim and N.~Kim, \emph{{1/4-BPS M-theory bubbles with SO(3) x
  SO(4) symmetry}},
  \href{http://dx.doi.org/10.1088/1126-6708/2007/08/050}{\emph{JHEP} {\bf 08}
  (2007) 050}, [\href{https://arxiv.org/abs/0706.2042}{{\tt 0706.2042}}].

\bibitem{Figueras:2007cn}
P.~Figueras, O.~A.~P. Mac~Conamhna and E.~O~Colgain, \emph{{Global geometry of
  the supersymmetric AdS(3)/CFT(2) correspondence in M-theory}},
  \href{http://dx.doi.org/10.1103/PhysRevD.76.046007}{\emph{Phys. Rev. D} {\bf
  76} (2007) 046007}, [\href{https://arxiv.org/abs/hep-th/0703275}{{\tt
  hep-th/0703275}}].

\bibitem{Donos:2008hd}
A.~Donos, J.~P. Gauntlett and J.~Sparks, \emph{{AdS(3) x (S**3 x S**3 x S**1)
  Solutions of Type IIB String Theory}},
  \href{http://dx.doi.org/10.1088/0264-9381/26/6/065009}{\emph{Class. Quant.
  Grav.} {\bf 26} (2009) 065009}, [\href{https://arxiv.org/abs/0810.1379}{{\tt
  0810.1379}}].

\bibitem{OColgain:2010wlk}
E.~O~Colgain, J.-B. Wu and H.~Yavartanoo, \emph{{Supersymmetric AdS3 X S2
  M-theory geometries with fluxes}},
  \href{http://dx.doi.org/10.1007/JHEP08(2010)114}{\emph{JHEP} {\bf 08} (2010)
  114}, [\href{https://arxiv.org/abs/1005.4527}{{\tt 1005.4527}}].

\bibitem{DHoker:2008lup}
E.~D'Hoker, J.~Estes, M.~Gutperle and D.~Krym, \emph{{Exact Half-BPS Flux
  Solutions in M-theory. I: Local Solutions}},
  \href{http://dx.doi.org/10.1088/1126-6708/2008/08/028}{\emph{JHEP} {\bf 08}
  (2008) 028}, [\href{https://arxiv.org/abs/0806.0605}{{\tt 0806.0605}}].

\bibitem{Estes:2012vm}
J.~Estes, R.~Feldman and D.~Krym, \emph{{Exact half-BPS flux solutions in $M$
  theory with D(2,1;$c^\prime$;0)$^2$ symmetry: Local solutions}},
  \href{http://dx.doi.org/10.1103/PhysRevD.87.046008}{\emph{Phys. Rev. D} {\bf
  87} (2013) 046008}, [\href{https://arxiv.org/abs/1209.1845}{{\tt
  1209.1845}}].

\bibitem{Bachas:2013vza}
C.~Bachas, E.~D'Hoker, J.~Estes and D.~Krym, \emph{{M-theory Solutions
  Invariant under $D(2,1;\gamma) \oplus D(2,1;\gamma)$}},
  \href{http://dx.doi.org/10.1002/prop.201300039}{\emph{Fortsch. Phys.} {\bf
  62} (2014) 207--254}, [\href{https://arxiv.org/abs/1312.5477}{{\tt
  1312.5477}}].

\bibitem{Benini:2013cda}
F.~Benini and N.~Bobev, \emph{{Two-dimensional SCFTs from wrapped branes and
  c-extremization}},
  \href{http://dx.doi.org/10.1007/JHEP06(2013)005}{\emph{JHEP} {\bf 06} (2013)
  005}, [\href{https://arxiv.org/abs/1302.4451}{{\tt 1302.4451}}].

\bibitem{Jeong:2014iva}
J.~Jeong, E.~\'O~Colg\'ain and K.~Yoshida, \emph{{SUSY properties of warped
  $AdS_3$}}, \href{http://dx.doi.org/10.1007/JHEP06(2014)036}{\emph{JHEP} {\bf
  06} (2014) 036}, [\href{https://arxiv.org/abs/1402.3807}{{\tt 1402.3807}}].

\bibitem{Lozano:2015bra}
Y.~Lozano, N.~T. Macpherson, J.~Montero and E.~O. Colg\'ain, \emph{{New $AdS_3
  \times S^2$ T-duals with $ \mathcal{N}=\left(0,4\right) $ supersymmetry}},
  \href{http://dx.doi.org/10.1007/JHEP08(2015)121}{\emph{JHEP} {\bf 08} (2015)
  121}, [\href{https://arxiv.org/abs/1507.02659}{{\tt 1507.02659}}].

\bibitem{Benini:2015bwz}
F.~Benini, N.~Bobev and P.~M. Crichigno, \emph{{Two-dimensional SCFTs from
  D3-branes}}, \href{http://dx.doi.org/10.1007/JHEP07(2016)020}{\emph{JHEP}
  {\bf 07} (2016) 020}, [\href{https://arxiv.org/abs/1511.09462}{{\tt
  1511.09462}}].

\bibitem{Kelekci:2016uqv}
O.~Kelekci, Y.~Lozano, J.~Montero, E.~O. Colg\'ain and M.~Park, \emph{{Large
  superconformal near-horizons from M-theory}},
  \href{http://dx.doi.org/10.1103/PhysRevD.93.086010}{\emph{Phys. Rev. D} {\bf
  93} (2016) 086010}, [\href{https://arxiv.org/abs/1602.02802}{{\tt
  1602.02802}}].

\bibitem{Eberhardt:2017uup}
L.~Eberhardt, \emph{{Supersymmetric AdS$_{3}$ supergravity backgrounds and
  holography}}, \href{http://dx.doi.org/10.1007/JHEP02(2018)087}{\emph{JHEP}
  {\bf 02} (2018) 087}, [\href{https://arxiv.org/abs/1710.09826}{{\tt
  1710.09826}}].

\bibitem{Dibitetto:2018iar}
G.~Dibitetto and N.~Petri, \emph{{Surface defects in the D4 $-$ D8 brane
  system}}, \href{http://dx.doi.org/10.1007/JHEP01(2019)193}{\emph{JHEP} {\bf
  01} (2019) 193}, [\href{https://arxiv.org/abs/1807.07768}{{\tt 1807.07768}}].

\bibitem{Dibitetto:2018ftj}
G.~Dibitetto, G.~Lo~Monaco, A.~Passias, N.~Petri and A.~Tomasiello,
  \emph{{AdS$_3$ Solutions with Exceptional Supersymmetry}},
  \href{http://dx.doi.org/10.1002/prop.201800060}{\emph{Fortsch. Phys.} {\bf
  66} (2018) 1800060}, [\href{https://arxiv.org/abs/1807.06602}{{\tt
  1807.06602}}].

\bibitem{Macpherson:2018mif}
N.~T. Macpherson, \emph{{Type II solutions on AdS$_{3} \times$ S$^{3} \times$
  S$^{3}$ with large superconformal symmetry}},
  \href{http://dx.doi.org/10.1007/JHEP05(2019)089}{\emph{JHEP} {\bf 05} (2019)
  089}, [\href{https://arxiv.org/abs/1812.10172}{{\tt 1812.10172}}].

\bibitem{Legramandi:2019xqd}
A.~Legramandi and N.~T. Macpherson, \emph{{AdS$_3$ solutions with from
  $\mathcal{N}=(3,0)$ from S$^3\times$S$^3$ fibrations}},
  \href{http://dx.doi.org/10.1002/prop.202000014}{\emph{Fortsch. Phys.} {\bf
  68} (2020) 2000014}, [\href{https://arxiv.org/abs/1912.10509}{{\tt
  1912.10509}}].

\bibitem{Lozano:2019jza}
Y.~Lozano, N.~T. Macpherson, C.~Nunez and A.~Ramirez, \emph{{1/4 BPS solutions
  and the AdS$_3$/CFT$_2$ correspondence}},
  \href{http://dx.doi.org/10.1103/PhysRevD.101.026014}{\emph{Phys. Rev. D} {\bf
  101} (2020) 026014}, [\href{https://arxiv.org/abs/1909.09636}{{\tt
  1909.09636}}].

\bibitem{Lozano:2019zvg}
Y.~Lozano, N.~T. Macpherson, C.~Nunez and A.~Ramirez, \emph{{Two dimensional
  ${\cal N}=(0,4)$ quivers dual to AdS$_3$ solutions in massive IIA}},
  \href{http://dx.doi.org/10.1007/JHEP01(2020)140}{\emph{JHEP} {\bf 01} (2020)
  140}, [\href{https://arxiv.org/abs/1909.10510}{{\tt 1909.10510}}].

\bibitem{Lozano:2019ywa}
Y.~Lozano, N.~T. Macpherson, C.~Nunez and A.~Ramirez, \emph{{AdS$_3$ solutions
  in massive IIA, defect CFTs and T-duality}},
  \href{http://dx.doi.org/10.1007/JHEP12(2019)013}{\emph{JHEP} {\bf 12} (2019)
  013}, [\href{https://arxiv.org/abs/1909.11669}{{\tt 1909.11669}}].

\bibitem{Couzens:2019mkh}
C.~Couzens, H.~het Lam and K.~Mayer, \emph{{Twisted $ \mathcal{N} $ = 1 SCFTs
  and their AdS$_{3}$ duals}},
  \href{http://dx.doi.org/10.1007/JHEP03(2020)032}{\emph{JHEP} {\bf 03} (2020)
  032}, [\href{https://arxiv.org/abs/1912.07605}{{\tt 1912.07605}}].

\bibitem{Couzens:2019iog}
C.~Couzens, \emph{{$ \mathcal{N} $ = (0, 2) AdS$_{3}$ solutions of type IIB and
  F-theory with generic fluxes}},
  \href{http://dx.doi.org/10.1007/JHEP04(2021)038}{\emph{JHEP} {\bf 04} (2021)
  038}, [\href{https://arxiv.org/abs/1911.04439}{{\tt 1911.04439}}].

\bibitem{Passias:2019rga}
A.~Passias and D.~Prins, \emph{{On AdS$_3$ solutions of Type IIB}},
  \href{http://dx.doi.org/10.1007/JHEP05(2020)048}{\emph{JHEP} {\bf 05} (2020)
  048}, [\href{https://arxiv.org/abs/1910.06326}{{\tt 1910.06326}}].

\bibitem{Filippas:2019ihy}
K.~Filippas, \emph{{Non-integrability on AdS$_{3}$ supergravity backgrounds}},
  \href{http://dx.doi.org/10.1007/JHEP02(2020)027}{\emph{JHEP} {\bf 02} (2020)
  027}, [\href{https://arxiv.org/abs/1910.12981}{{\tt 1910.12981}}].

\bibitem{Speziali:2019uzn}
S.~Speziali, \emph{{Spin 2 fluctuations in 1/4 BPS AdS$_3$/CFT$_2$}},
  \href{http://dx.doi.org/10.1007/JHEP03(2020)079}{\emph{JHEP} {\bf 03} (2020)
  079}, [\href{https://arxiv.org/abs/1910.14390}{{\tt 1910.14390}}].

\bibitem{Lozano:2020bxo}
Y.~Lozano, C.~Nunez, A.~Ramirez and S.~Speziali, \emph{{$M$-strings and AdS$_3$
  solutions to M-theory with small $\mathcal{N}=(0,4)$ supersymmetry}},
  \href{http://dx.doi.org/10.1007/JHEP08(2020)118}{\emph{JHEP} {\bf 08} (2020)
  118}, [\href{https://arxiv.org/abs/2005.06561}{{\tt 2005.06561}}].

\bibitem{Farakos:2020phe}
F.~Farakos, G.~Tringas and T.~Van~Riet, \emph{{No-scale and scale-separated
  flux vacua from IIA on G2 orientifolds}},
  \href{http://dx.doi.org/10.1140/epjc/s10052-020-8247-5}{\emph{Eur. Phys. J.
  C} {\bf 80} (2020) 659}, [\href{https://arxiv.org/abs/2005.05246}{{\tt
  2005.05246}}].

\bibitem{Rigatos:2020igd}
K.~S. Rigatos, \emph{{Non-integrability in AdS$_{3}$ vacua}},
  \href{http://dx.doi.org/10.1007/JHEP02(2021)032}{\emph{JHEP} {\bf 02} (2021)
  032}, [\href{https://arxiv.org/abs/2011.08224}{{\tt 2011.08224}}].

\bibitem{Faedo:2020nol}
F.~Faedo, Y.~Lozano and N.~Petri, \emph{{Searching for surface defect CFTs
  within AdS$_3$}},
  \href{http://dx.doi.org/10.1007/JHEP11(2020)052}{\emph{JHEP} {\bf 11} (2020)
  052}, [\href{https://arxiv.org/abs/2007.16167}{{\tt 2007.16167}}].

\bibitem{Dibitetto:2020bsh}
G.~Dibitetto and N.~Petri, \emph{{AdS$_{3}$ from M-branes at conical
  singularities}}, \href{http://dx.doi.org/10.1007/JHEP01(2021)129}{\emph{JHEP}
  {\bf 01} (2021) 129}, [\href{https://arxiv.org/abs/2010.12323}{{\tt
  2010.12323}}].

\bibitem{Filippas:2020qku}
K.~Filippas, \emph{{Holography for 2D $\mathcal{N}=(0,4)$ quantum field
  theory}}, \href{http://dx.doi.org/10.1103/PhysRevD.103.086003}{\emph{Phys.
  Rev. D} {\bf 103} (2021) 086003},
  [\href{https://arxiv.org/abs/2008.00314}{{\tt 2008.00314}}].

\bibitem{Passias:2020ubv}
A.~Passias and D.~Prins, \emph{{On supersymmetric AdS$_3$ solutions of Type
  II}}, \href{http://dx.doi.org/10.1007/JHEP08(2021)168}{\emph{JHEP} {\bf 08}
  (2021) 168}, [\href{https://arxiv.org/abs/2011.00008}{{\tt 2011.00008}}].

\bibitem{Faedo:2020lyw}
F.~Faedo, Y.~Lozano and N.~Petri, \emph{{New $\mathcal{N}=(0,4)$ AdS$_3$
  near-horizons in Type IIB}},
  \href{http://dx.doi.org/10.1007/JHEP04(2021)028}{\emph{JHEP} {\bf 04} (2021)
  028}, [\href{https://arxiv.org/abs/2012.07148}{{\tt 2012.07148}}].

\bibitem{Eloy:2020uix}
C.~Eloy, \emph{{Kaluza-Klein spectrometry for ${\rm AdS_{3}}$ vacua}},
  \href{http://dx.doi.org/10.21468/SciPostPhys.10.6.131}{\emph{SciPost Phys.}
  {\bf 10} (2021) 131}, [\href{https://arxiv.org/abs/2011.11658}{{\tt
  2011.11658}}].

\bibitem{Legramandi:2020txf}
A.~Legramandi, G.~Lo~Monaco and N.~T. Macpherson, \emph{{All
  $\mathcal{N}=(8,0)$ AdS$_3$ solutions in 10 and 11 dimensions}},
  \href{http://dx.doi.org/10.1007/JHEP05(2021)263}{\emph{JHEP} {\bf 05} (2021)
  263}, [\href{https://arxiv.org/abs/2012.10507}{{\tt 2012.10507}}].

\bibitem{Zacarias:2021pfz}
S.~Zacarias, \emph{{Marginal deformations of a class of AdS$_{3} \mathcal{N} $
  = (0, 4) holographic backgrounds}},
  \href{http://dx.doi.org/10.1007/JHEP06(2021)017}{\emph{JHEP} {\bf 06} (2021)
  017}, [\href{https://arxiv.org/abs/2102.05681}{{\tt 2102.05681}}].

\bibitem{Emelin:2021gzx}
M.~Emelin, F.~Farakos and G.~Tringas, \emph{{Three-dimensional flux vacua from
  IIB on co-calibrated G2 orientifolds}},
  \href{http://dx.doi.org/10.1140/epjc/s10052-021-09261-y}{\emph{Eur. Phys. J.
  C} {\bf 81} (2021) 456}, [\href{https://arxiv.org/abs/2103.03282}{{\tt
  2103.03282}}].

\bibitem{Couzens:2021tnv}
C.~Couzens, N.~T. Macpherson and A.~Passias, \emph{{${\cal N}=(2,2)$ AdS$_3$
  from D3-branes wrapped on Riemann surfaces}},
  \href{https://arxiv.org/abs/2107.13562}{{\tt 2107.13562}}.

\bibitem{Suh:2021ifj}
M.~Suh, \emph{{D3-branes and M5-branes wrapped on a topological disc}},
  \href{https://arxiv.org/abs/2108.01105}{{\tt 2108.01105}}.

\bibitem{Boido:2021szx}
A.~Boido, J.~M.~P. Ipi\~na and J.~Sparks, \emph{{Twisted D3-brane and M5-brane
  compactifications from multi-charge spindles}},
  \href{http://dx.doi.org/10.1007/JHEP07(2021)222}{\emph{JHEP} {\bf 07} (2021)
  222}, [\href{https://arxiv.org/abs/2104.13287}{{\tt 2104.13287}}].

\bibitem{Ferrero:2020laf}
P.~Ferrero, J.~P. Gauntlett, J.~M. P\'erez Ipi\~na, D.~Martelli and J.~Sparks,
  \emph{{D3-Branes Wrapped on a Spindle}},
  \href{http://dx.doi.org/10.1103/PhysRevLett.126.111601}{\emph{Phys. Rev.
  Lett.} {\bf 126} (2021) 111601},
  [\href{https://arxiv.org/abs/2011.10579}{{\tt 2011.10579}}].

\bibitem{Hosseini:2021fge}
S.~M. Hosseini, K.~Hristov and A.~Zaffaroni, \emph{{Rotating multi-charge
  spindles and their microstates}},
  \href{http://dx.doi.org/10.1007/JHEP07(2021)182}{\emph{JHEP} {\bf 07} (2021)
  182}, [\href{https://arxiv.org/abs/2104.11249}{{\tt 2104.11249}}].

\bibitem{Bernamonti:2007bu}
A.~Bernamonti, M.~M. Caldarelli, D.~Klemm, R.~Olea, C.~Sieg and E.~Zorzan,
  \emph{{Black strings in AdS(5)}},
  \href{http://dx.doi.org/10.1088/1126-6708/2008/01/061}{\emph{JHEP} {\bf 01}
  (2008) 061}, [\href{https://arxiv.org/abs/0708.2402}{{\tt 0708.2402}}].

\bibitem{Hosseini:2016cyf}
S.~M. Hosseini, A.~Nedelin and A.~Zaffaroni, \emph{{The Cardy limit of the
  topologically twisted index and black strings in AdS$_{5}$}},
  \href{http://dx.doi.org/10.1007/JHEP04(2017)014}{\emph{JHEP} {\bf 04} (2017)
  014}, [\href{https://arxiv.org/abs/1611.09374}{{\tt 1611.09374}}].

\bibitem{Azzola:2018sld}
M.~Azzola, D.~Klemm and M.~Rabbiosi, \emph{{AdS$_5$ black strings in the stu
  model of FI-gauged $N=2$ supergravity}},
  \href{http://dx.doi.org/10.1007/JHEP10(2018)080}{\emph{JHEP} {\bf 10} (2018)
  080}, [\href{https://arxiv.org/abs/1803.03570}{{\tt 1803.03570}}].

\bibitem{Hosseini:2019lkt}
S.~M. Hosseini, K.~Hristov and A.~Zaffaroni, \emph{{Microstates of rotating
  AdS$_{5}$ strings}},
  \href{http://dx.doi.org/10.1007/JHEP11(2019)090}{\emph{JHEP} {\bf 11} (2019)
  090}, [\href{https://arxiv.org/abs/1909.08000}{{\tt 1909.08000}}].

\bibitem{Hosseini:2020vgl}
S.~M. Hosseini, K.~Hristov, Y.~Tachikawa and A.~Zaffaroni, \emph{{Anomalies,
  Black strings and the charged Cardy formula}},
  \href{http://dx.doi.org/10.1007/JHEP09(2020)167}{\emph{JHEP} {\bf 09} (2020)
  167}, [\href{https://arxiv.org/abs/2006.08629}{{\tt 2006.08629}}].

\bibitem{Cvetic:2000mh}
M.~Cvetic, H.~Lu and C.~Pope, \emph{{Brane resolution through transgression}},
  \href{http://dx.doi.org/10.1016/S0550-3213(01)00050-5}{\emph{Nucl. Phys. B}
  {\bf 600} (2001) 103--132}, [\href{https://arxiv.org/abs/hep-th/0011023}{{\tt
  hep-th/0011023}}].

\bibitem{Brown:1986nw}
J.~D. Brown and M.~Henneaux, \emph{{Central Charges in the Canonical
  Realization of Asymptotic Symmetries: An Example from Three-Dimensional
  Gravity}}, \href{http://dx.doi.org/10.1007/BF01211590}{\emph{Commun. Math.
  Phys.} {\bf 104} (1986) 207--226}.

\bibitem{Hanany:1997sa}
A.~Hanany and A.~Zaffaroni, \emph{{Chiral symmetry from type IIA branes}},
  \href{http://dx.doi.org/10.1016/S0550-3213(97)00595-6}{\emph{Nucl. Phys. B}
  {\bf 509} (1998) 145--168}, [\href{https://arxiv.org/abs/hep-th/9706047}{{\tt
  hep-th/9706047}}].

\bibitem{Cremonesi:2015bld}
S.~Cremonesi and A.~Tomasiello, \emph{{6d holographic anomaly match as a
  continuum limit}},
  \href{http://dx.doi.org/10.1007/JHEP05(2016)031}{\emph{JHEP} {\bf 05} (2016)
  031}, [\href{https://arxiv.org/abs/1512.02225}{{\tt 1512.02225}}].

\bibitem{Lozano:2020txg}
Y.~Lozano, C.~Nunez, A.~Ramirez and S.~Speziali, \emph{{New AdS$_{2}$
  backgrounds and $ \mathcal{N} $ = 4 conformal quantum mechanics}},
  \href{http://dx.doi.org/10.1007/JHEP03(2021)277}{\emph{JHEP} {\bf 03} (2021)
  277}, [\href{https://arxiv.org/abs/2011.00005}{{\tt 2011.00005}}].

\bibitem{Lozano:2020sae}
Y.~Lozano, C.~Nunez, A.~Ramirez and S.~Speziali, \emph{{AdS$_{2}$ duals to ADHM
  quivers with Wilson lines}},
  \href{http://dx.doi.org/10.1007/JHEP03(2021)145}{\emph{JHEP} {\bf 03} (2021)
  145}, [\href{https://arxiv.org/abs/2011.13932}{{\tt 2011.13932}}].

\bibitem{Lozano:2021rmk}
Y.~Lozano, C.~Nunez and A.~Ramirez, \emph{{$\text{AdS}_2\times \text{S}^2\times
  \text{CY}_2$ solutions in Type IIB with 8 supersymmetries}},
  \href{http://dx.doi.org/10.1007/JHEP04(2021)110}{\emph{JHEP} {\bf 04} (2021)
  110}, [\href{https://arxiv.org/abs/2101.04682}{{\tt 2101.04682}}].

\bibitem{Lozano:2021fkk}
Y.~Lozano, N.~Petri and C.~Risco, \emph{{New AdS$_2$ supergravity duals of 4d
  SCFTs with defects}},  \href{https://arxiv.org/abs/2107.12277}{{\tt
  2107.12277}}.

\bibitem{Couzens:2020jgx}
C.~Couzens, E.~Marcus, K.~Stemerdink and D.~van~de Heisteeg, \emph{{The
  near-horizon geometry of supersymmetric rotating AdS$_{4}$ black holes in
  M-theory}}, \href{http://dx.doi.org/10.1007/JHEP05(2021)194}{\emph{JHEP} {\bf
  05} (2021) 194}, [\href{https://arxiv.org/abs/2011.07071}{{\tt 2011.07071}}].

\bibitem{Tong:2014yna}
D.~Tong, \emph{{The holographic dual of $AdS_{3} \times S^{3} \times S^{3}
  \times S^{1}$}}, \href{http://dx.doi.org/10.1007/JHEP04(2014)193}{\emph{JHEP}
  {\bf 04} (2014) 193}, [\href{https://arxiv.org/abs/1402.5135}{{\tt
  1402.5135}}].

\bibitem{Hanany:1996ie}
A.~Hanany and E.~Witten, \emph{{Type IIB superstrings, BPS monopoles, and
  three-dimensional gauge dynamics}},
  \href{http://dx.doi.org/10.1016/S0550-3213(97)00157-0}{\emph{Nucl. Phys. B}
  {\bf 492} (1997) 152--190}, [\href{https://arxiv.org/abs/hep-th/9611230}{{\tt
  hep-th/9611230}}].

\bibitem{Banks:1997zs}
T.~Banks, N.~Seiberg and E.~Silverstein, \emph{{Zero and one-dimensional probes
  with N=8 supersymmetry}},
  \href{http://dx.doi.org/10.1016/S0370-2693(97)00366-3}{\emph{Phys. Lett. B}
  {\bf 401} (1997) 30--37}, [\href{https://arxiv.org/abs/hep-th/9703052}{{\tt
  hep-th/9703052}}].

\bibitem{Kraus:2005zm}
P.~Kraus and F.~Larsen, \emph{{Holographic gravitational anomalies}},
  \href{http://dx.doi.org/10.1088/1126-6708/2006/01/022}{\emph{JHEP} {\bf 01}
  (2006) 022}, [\href{https://arxiv.org/abs/hep-th/0508218}{{\tt
  hep-th/0508218}}].

\end{thebibliography}\endgroup

\end{document}